\documentclass[iop]{emulateapj}
\usepackage{amssymb,amsmath,amsfonts,graphicx,color,epsf}

\newcommand{\beq}{\begin{equation}}
\newcommand{\eeq}{\end{equation}}

\renewcommand{\vec}[1]{\mathbf{#1}}

\renewcommand{\(}{\left(}
\renewcommand{\)}{\right)}

\begin{document}

\title{Statistical Analysis of Current Sheets in Three-Dimensional Magnetohydrodynamic Turbulence}

\author{Vladimir Zhdankin}
\email{zhdankin@wisc.edu}
\affiliation{Department of Physics, University of Wisconsin, Madison, WI 53706} 

\author{Dmitri A. Uzdensky}
\email{uzdensky@colorado.edu}
\affiliation{Center for Integrated Plasma Studies, Physics Department, UCB-390, University of Colorado, Boulder, CO 80309} 

\author{Jean C. Perez}
\email{jcperez@wisc.edu}
\affiliation{Space Science Center, University of New Hampshire, Durham, NH 03824, USA} 

\author{Stanislav Boldyrev}
\email{boldyrev@wisc.edu}
\affiliation{Department of Physics, University of Wisconsin, Madison, WI 53706, USA}

\date{\today}


\begin{abstract}
We develop a framework for studying the statistical properties of current sheets in numerical simulations of 3D magnetohydrodynamic (MHD) turbulence. We describe an algorithm that identifies current sheets in a simulation snapshot and then determines their geometrical properties (including length, width, and thickness) and intensities (peak current density and total energy dissipation rate). We then apply this procedure to simulations of reduced MHD turbulence and perform a statistical analysis on the obtained population of current sheets. We evaluate the role of reconnection by separately studying the populations of current sheets which contain magnetic X-points and those which do not. We find that the statistical properties of the two populations are different in general. We compare the scaling of these properties to phenomenological predictions obtained for the inertial range of MHD turbulence. Finally, we test whether the reconnecting current sheets are consistent with the Sweet-Parker model.

\end{abstract}


\maketitle


\section{Introduction}
\label{sec-intro}

Magnetic reconnection and turbulence are two of the most important and well studied processes in modern plasma physics and plasma astrophysics. Both of them have been recognized as critical to understanding many laboratory, space, and astrophysical phenomena and have been the subjects of extensive research in the past few decades. However, in many real natural systems, these processes do not occur in isolation, independent of each other. Real systems are often complex; they usually involve many interacting components and many processes acting simultaneously. It is therefore clear that, in order to understand how reconnection and turbulence operate in such systems, one needs to understand how they interact and affect each other.

In principle, one can define two sets of questions: (1) how a relatively small-scale externally superimposed or internally-generated turbulence affects reconnection of a large-scale magnetic field; and (2) how reconnection at the bottom of the turbulent cascade affects/controls energy dissipation in magnetohydrodynamic (MHD) turbulence. Despite the obvious importance of both of these fundamental physics questions, so far relatively little research has been done on them, compared with the number of studies dealing with either turbulence or reconnection separately. 

On the first topic, we note that, even though the potential importance of externally imposed magnetohydrodynamic turbulence in reconnection has been recognized early on \cite[e.g.,][]{matthaeus1986, lazarian1999, kim2001}, most of the traditional studies of reconnection have assumed a completely smooth and laminar background/initial state. Only relatively recently, with the advent of more powerful numerical simulations, has this topic received a proper attention \citep{kowal2009,loureiro2009}. In addition to these studies, which focused on the effect of an externally driven turbulence on reconnection, one can also ask whether and how a reconnection process is affected by internally-generated MHD turbulence, produced as a by-product of the reconnection process itself \cite[e.g.,][]{strauss1988}. 
This question has also received a lot of attention in recent years, largely in connection with the transition of a new dynamic, non-stationary regime of fast resistive-MHD reconnection mediated by the secondary tearing/plasmoid instability \citep{loureiro2007}. This so-called plasmoid reconnection regime has been studied both analytically \citep{loureiro2007, uzdensky2010, baalrud2012, loureiro2013} and numerically \citep{lapenta2008, samtaney2009,bhattacharjee2009,cassak2009,huang2010,loureiro2012} and can be seen as a special form of turbulent reconnection. 

As for the second topic --- turbulent energy dissipation via magnetic reconnection at the bottom  of the turbulent cascade --- the number of papers published on it so far has been much less. The most notable was a recent series of statistical studies by Servidio et al. \citep{servidio2009,servidio2010,servidio2011,servidio2011b}.  There has also been a more general investigation of energy dissipation in intermittent current and vorticity structures performed by \cite{uritsky2010}. We discuss these papers in more detail and compare them with our analysis in the last section of this paper.

In addition to the obvious fundamental-physics motivation for investigating reconnection in turbulence, there are also practical applications to various astrophysical environments.  At issue here is the larger question of the intermittency of energy dissipation and hence of plasma heating in MHD turbulence. This question becomes especially important in situations where there is strong prompt radiative cooling that may evacuate the dissipated energy from localized dissipation sites (hot spots) faster than it can be redistributed uniformly across the plasma by turbulent diffusion or thermal conduction.  This can greatly alter the thermodynamics of such systems; in particular, lead to a highly inhomogeneous temperature structure.  An important astrophysical example of a system where this may happen is coronae of accretion disks around black holes accreting at a near-Eddington limit, e.g., in quasars, where $\sim 100$~keV electrons are subject to very powerful inverse-Compton  cooling \citep{goodman2008}. However, these issues may also be important in  various other high-energy astrophysical systems, such as accretion disks and flows, hot X-ray gas in some galaxy clusters, etc.  Furthermore, even in low-energy-density astrophysical systems, where prompt radiative cooling is not important, the issue of intermittency of turbulent  energy dissipation may still be of interest for another reason.  Namely, it determines the spatial distribution and coherence lengths of electric fields, and thus may affect the effectiveness of nonthermal particle acceleration in collisionless or weakly collisional plasmas in, e.g., radiatively-inefficient accretion flows, galaxy clusters, and the solar wind. 

Another application is the dissipation and intermittency of the solar wind.  Current sheets are associated with magnetic discontinuities \citep{greco2010}, which are observed $\text{\it{in-situ}}$ by spacecraft in the solar wind and have been studied in a number of works recently. The origin of the discontinuities is not well understood and attracts considerable interest. Two possible explanations are discussed in the literature. One possibility is that the discontinuities are generated in the solar corona and then advected by the solar wind \citep{borovsky2008,miao2011}. The other possibility is that they are dynamically generated in the solar wind, for example, due to inherent magnetic plasma turbulence \citep{li_etal11,borovsky10,zhdankin2012}. The investigation of current sheets formed in turbulence, and their relationship to observable magnetic discontinuities, may therefore be useful for distinguishing between these two possibilities. 

The present paper is devoted to a detailed study of the intermittency of magnetic energy dissipation in magnetohydrodynamic (MHD) turbulence and, more specifically, to assessing the role  of small-scale magnetic reconnection events at the bottom of the turbulent cascade in the overall turbulent heating process.  For simplicity, in this study we focus only on small-scale magnetic structures and associated ohmic heating, leaving viscous dissipation of kinetic energy to a future study. Since ohmic heating per unit volume is equal~$\eta j^2$ and we assume constant resistivity $\eta$, the main sites of magnetic energy dissipation are the regions of concentration of the  current density~$j$.  In three-dimensional MHD turbulence such regions are two-dimensional current sheets, which may or may not be associated with magnetic reconnection (i.e., with a change of magnetic field topology, marked by the presence of magnetic X-points or X-lines). 

Correspondingly, the main focus of the present paper is on a statistical study of intense current sheets and their associated heating rates in two- and three-dimensional (2D and 3D) MHD turbulence.  In this study we are interested in looking at current sheets as a population and want to address statistical questions such as:  
How important are intense current sheets in the overall energy dissipation?  
What is the distribution of the current sheets in intensity (i.e., the peak values of the volumetric current density, $j_{\rm max}$) and in integrated energy dissipation/heating rates?  
What is the distribution of their geometric properties such as lengths, widths, and thicknesses? 
How are all these quantities correlated with each other?  
Are there substantial statistical differences between current sheets that are associated with reconnection events (containing an X-point) and those that are not?  
What is the role of discrete reconnection events in the overall magnetic energy dissipation and in its intermittency properties?  

To address these questions, we develop a set of numerical procedures and tools for identifying current sheets in simulation data and quantiatively characterizing them in terms of their peak current density, dimensions, overall magnetic dissipation rate, etc. Our algorithm allows us to distinguish between reconnecting current sheets (those with X-points) and non-reconnecting ones (without X-points).
We then apply these tools to post-process our existing high-resolution (up to $1024^3$ grid points) numerical simulations of driven MHD turbulence.  

As a result of this study, we are able to show that a large number of the current sheets do not contain reconnection sites, and likewise, many reconnection sites do not reside in intense current sheets.  However, the current sheets that do contain reconnection sites tend to be larger and more intense, and their properties show more robust scaling relationships than current sheets without reconnection sites. We find that the scalings differ from what is expected from phenomenological modeling of the inertial range of MHD turbulence, apparently reflecting the fact that current sheet formation also depends on the processes in the dissipation range of the energy spectrum. 
A comparison for reconnecting current sheets of measured thickness with the thickness expected from the Sweet-Parker model shows reasonable agreement.

This paper is organized as follows. In Section \ref{sec-dissipation-scale}, we discuss phenomenological estimates for the dissipation scale in MHD turbulence.  In Section~\ref{sec-tools}, we describe our numerical MHD simulations (\S~\ref{subsec-mhd-sims}) and the algorithm we developed to identify and characterize current sheets (\S~\ref{subsec-algorithms}).  In Section~\ref{sec-results}, we describe the results of our statistical analysis of the current sheets and compare to the Sweet-Parker theory.  In Section \ref{sec-conclusions}, we compare our results to similar studies, discuss the potential implications for various space and astrophysical systems, and outline future extensions of this work.  Finally, in Section \ref{sec-future} we summarize our conclusions.


\section{Fiducial dissipation scale of MHD turbulence}
\label{sec-dissipation-scale}

The incompressible MHD equations take the form
\begin{eqnarray}
 \label{mhd-elsasser}
  \left( \frac{\partial}{\partial t}\mp\vec V_A\cdot\nabla \right)
  \vec z^\pm+\left(\vec z^\mp\cdot\nabla\right)\vec z^\pm &=& -\nabla
  P +\nu\nabla^2 \vec z^{\pm}+\vec f^\pm, \nonumber \\ \nabla \cdot
  {\vec z}^{\pm}&=&0
  \end{eqnarray}
where $\vec z^\pm=\vec v\pm\vec b$ are the Els\"asser variables, $\vec v$ is the fluctuating plasma velocity, $\vec b$ is the fluctuating magnetic field (in units of the Alfv\'en velocity, $\vec{V_A}={\bf
  B}_0/\sqrt{4\pi\rho_0}$, based upon the uniform background magnetic field $\vec{B_0}$ and plasma density~$\rho_0$), $P=(p/\rho_0+b^2/2)$, $p$ is the plasma pressure, $\nu$ is the fluid viscosity (which, for simplicity, we have taken to be equal to the magnetic diffusivity), and $\vec f^\pm$ represent forces that drive the turbulence at large scales. It can be shown that in the limit of small amplitude fluctuations, and in the absence of forcing and dissipation, the system describes non-interacting linear Alfv\'en waves with the  dispersion relation $\omega^\pm(\vec k)=\pm k_\| V_A$. 
The incompressibility condition requires that these waves be transverse.  
Typically they are decomposed into shear Alfv\'en waves (with polarizations 
perpendicular to both $\vec{B_0}$ and to the wave-vector $\vec{k}$) and 
pseudo-Alfv\'en waves (with polarizations in the plane of $\vec{B_0}$ and $\vec{k}$ 
and perpendicular to~$\vec{k}$).

In order to set the stage for the subsequent statistical study of intense intermittent dissipative structures in MHD turbulence, in this section we discuss the phenomenological  estimates for the dissipation scale and, more generally, for the typical turbulent dissipative structures.  These estimates are based on the analysis of the scaling relationships presented in \citep{boldyrev2005,boldyrev2006} that characterize the inertial range of the turbulence  and on evaluating the conditions when dissipative (e.g., resistive) terms become important.  Importantly, these estimates are done without taking into account intermittency of MHD turbulence and thus can serve as the simplest baseline against which intermittent dissipative structures,  which constitute the main focus of this paper, can be compared. 

According to the phenomenological picture of MHD turbulence discussed by \cite{goldreich1995} and \cite{boldyrev2005,boldyrev2006}, the typical structures within the inertial range are highly 3D-anisotropic.  
The degree of anisotropy of a turbulent eddy [which is related to the scales over which the typical (rms) fluctuations of velocity and magnetic field are correlated in different directions] increases as one goes down to smaller and smaller scales.  As argued by \cite{boldyrev2005,boldyrev2006}, it is characterized by the following relationships between the typical eddy scales in different directions: 
\begin{eqnarray}
\xi &\propto & \lambda^{3/4} \, , \label{widtheq} \\
l &\propto & \xi^{2/3} \propto \lambda^{1/2} \, . \label{lengtheq}
\end{eqnarray}
Here, $l$ is the scale of a given structure in the field-parallel direction, $\xi$ is the typical scale in the field-perpendicular direction {\em along} the fluctuating velocity and magnetic fields $\tilde{\bf v}_\lambda$ and $\tilde{\bf b}_\lambda$, and finally, $\lambda$ is the typical scale in the field-perpendicular direction but across $\tilde{\bf v}_\lambda$ and~$\tilde{\bf b}_\lambda$.\footnote{Note that the relationship $l \propto \xi^{2/3}$ would correspond to the Goldreich-Shridhar (1995) theory, although the scaling of the fluctuating field $b_\lambda$ and $v_\lambda$ is different in the phenomenology of \cite{boldyrev2006}, which we consider here and which agrees better with the numerical data.} 
Thus, we see that indeed, the turbulence is highly anisotropic, $l \gg \xi\gg \lambda$.

The typical velocity and magnetic field fluctuations in the inertial range scale as  \cite{boldyrev2005,boldyrev2006}
\beq
\tilde{v}_\lambda \propto \tilde{b}_\lambda \propto \lambda^{1/4} \, .
\eeq

Furthermore, from this we can obtain scalings for several other key quantities, e.g., the characteristic eddy turn-over time (this can be obtained from Goldreich-Shridhar critical balance argument): 
\beq
\tau_\lambda \propto l_\lambda/V_A \propto \xi_\lambda/\tilde{v}_\lambda \propto \lambda^{1/2} \, ,
\eeq
and the characteristic current density at a given scale: 
\beq
\tilde{j}_\lambda \propto {{\tilde{b}_\lambda}\over{\lambda}} \propto \lambda^{-3/4} \, .
\eeq

It should be noted that phenomenological models typically deal with an idealized picture which addresses the scaling of the fluctuating fields in the limit of very large Reynolds and magnetic Reynolds numbers. In reality these numbers are limited, and they are relatively small in numerical simulations (on the order of $10^4$). As was  shown in \citep{wang11,boldyrev_pz2012,boldyrev_pw2012}, in the case of limited inertial interval the scalings of magnetic and velocity fluctuations appear to be slightly different due to residual energy generated at large scales. The same effects are also observed in the solar wind turbulence \cite[e.g.,][]{boldyrev_pbp11}.  In our present discussion we do not take into account such effects, for two reasons.  First, the phenomenological theory is not developed well enough to address the scales near or inside the dissipation interval with the same certainty with which it addresses the inertial interval, and second, the statistical relations  measured in the present work are not necessarily related to the second-order correlation functions predicted by most phenomenological models.

Now let us turn to the discussion of the dissipation scale, which we will denote by $\lambda_\eta$. Similarly, all the quantities evaluated at the dissipation scale will be denoted by   the subscript~$\eta$ (assuming for definiteness/simplicity that viscosity~$\nu$ is equal to the magnetic diffusivity~$\eta$, i.e., that the magnetic Prandtl number is~1).  The scale $\lambda_\eta$ is defined as the scale at which the resistive diffusion time across it, $\tau_\eta=\lambda_\eta^2/\eta$,  is comparable to the corresponding eddy turn-over time, which gives 
\beq
\lambda_\eta \propto \eta^{2/3} \, , 
\eeq
and, correspondingly, 
\beq
\xi_\eta \propto \eta^{1/2} \, , \qquad l_\eta \propto \eta^{1/3} \, . 
\eeq
The other turbulent quantities at this scale are estimated as
\begin{eqnarray}
\tilde{v}_\eta &\sim&  \tilde{b}_\eta \equiv \tilde{b}_{\lambda_\eta} \propto \eta^{1/6}\, , \\
\tau_\eta &\propto & \eta^{1/3} \, , \\ 
\tilde{j}_\eta &\propto & \eta^{-1/2} \, . 
\label{eq-j_eta}
\end{eqnarray}

Finally, the characteristic energy dissipation rate in a single typical dissipative structure can be estimated (in~3D) as
\beq
\mathcal{E}_\eta \simeq \eta \tilde{j}_\eta^2 \, \lambda_\eta \xi_\eta l_\eta \propto \eta^{3/2} \, . 
\label{e-estimate}
\eeq

For reference, since most current state-of-the-art 3D numerical simulations of MHD turbulence have $\eta \sim 10^{-3}$ (in units normalized to the energy-containing length-scale and the rms velocity/magnetic field at that scale), the typical values of the above fiducial dissipation-scale parameters are: $\lambda_\eta \sim 0.01$, $\xi_\eta \sim 0.03$, $l_\eta \sim 0.1$, $\tilde{v}_\eta \sim \tilde{b}_\eta \sim 0.3$, $\tau_\eta \sim 0.1$, $\tilde{j}_\eta \sim 30$, and $\mathcal{E}_\eta \sim 3\times 10^{-5}$.

Next, it is interesting to note that the above estimates are consistent with the notion that such typical dissipative structures can be viewed as elementary Sweet-Parker \citep{sweet1958, parker1957} reconnection current sheets, in the sense that their lifetime $\tau_\eta$ is comparable to the Sweet-Parker (SP) reconnection time for the amount of flux equal to $\lambda_\eta \tilde{b}_\eta$, as one can easily see.  
The characteristic Lundquist number of these current sheets is $S_\eta \propto \tilde{b}_\eta \xi_\eta/\eta \propto \eta^{-1/3} \propto S_0^{1/3}$, where $S_0\propto \eta^{-1}$ is the global Lundquist number. Since in most modern simulations of 3D MHD turbulence this global Lundquist number does not exceed about 10,000, the Lundquist number of the typical SP current sheets at the dissipative scale is typically of order $S_\eta \sim 20$ or smaller, and correspondingly, their aspect ratio is not very large, $S_\eta^{1/2} \sim S_0^{1/6} \sim 4$.  Thus, it is not surprising that such current sheets are usually not recognized as thin SP current sheets in the simulation data. Their character as SP current sheets may have been more clearly visible if we had access to simulations with much larger~$S_0$ (not possible in the near future).

The estimates obtained in this section represent the phenomenological predictions for typical dissipation structures in 3D MHD turbulence. As we shall see in this paper, while they may be responsible for a significant fraction of the actual dissipation, they do not account for all of it; a sizable fraction of turbulent energy is dissipated in a few much more intense structures, and the main goal of this paper is to assess this contribution quantitatively.


\section{Numerical Procedures}
\label{sec-tools}


\subsection{Numerical MHD Simulations}
\label{subsec-mhd-sims}

For strong MHD turbulence, Goldreich \& Sridhar \citep{goldreich1995}
argued that the pseudo-Alfv\'en modes are dynamically irrelevant for
the turbulent cascade (since strong MHD turbulence is dominated by
fluctuations with $k_\perp \gg k_\|$, the polarization of the
pseudo-Alfv\'en fluctuations is almost parallel to the guide field and
they are therefore coupled only to field-parallel gradients, which are
small since $k_\| \ll k_{\perp}$). If one filters out the
pseudo-Alfv\'en modes by setting $\vec z^\pm_\|=0$, it can be shown
that the resulting system is equivalent to the Reduced MHD (RMHD) model:
\begin{eqnarray}
  \(\frac{\partial}{\partial t}\mp\vec V_A\cdot\nabla_\|\)\vec z^\pm
+\left(\vec z^\mp\cdot\nabla_\perp\right)\vec z^\pm = -\nabla_\perp P 
+\nu\nabla^2\vec z^\pm +\vec f_\perp^\pm \, . \nonumber \\
\label{rmhd-elsasser}
\end{eqnarray}  
We note that in RMHD the fluctuating fields have only two vector components, but that each depends on all three spatial coordinates. Moreover, because the $\vec z^\pm$ are assumed incompressible
($\nabla\cdot\vec z^\pm=0$), each field has only one degree of freedom; this is more commonly expressed in terms of stream functions in the more standard form of the RMHD equations~\citep{kadomtsev1974,strauss1976}.
This form is obtained if we introduce the axial component of the vector potential $\psi = -A_z$ and the stream function $\phi = \chi/B_0$, where $\chi$ is the scalar potential. The magnetic field and velocity are then recovered from $\psi$ and $\phi$ via 
 \begin{align}
 \boldsymbol{B} &= \boldsymbol{e}_z \times \nabla \psi + B_0 \boldsymbol{e}_z \nonumber \\
 \boldsymbol{u}_\bot &= \boldsymbol{e}_z \times \nabla \phi \, .
 \end{align}
 The reduced MHD equations are then written as:
 \begin{align}
 \frac{\partial \psi}{\partial t} - \boldsymbol{B} \cdot \nabla \phi - \frac{\eta}{\mu_0} \nabla^2 \psi &= 0 \nonumber \\
 \rho \frac{\partial \omega}{\partial t} + \rho \boldsymbol{u} \cdot \nabla \omega - \rho \nu \nabla^2 \omega &= \frac{1}{\mu_0} \boldsymbol{B} \cdot \nabla j
 \end{align}
 where $\omega = \nabla^2_\bot \phi$ and $j = \nabla^2_\bot \psi$.

We solve the RMHD equations (\ref{rmhd-elsasser}) in a periodic, rectangular domain
with dimensions $L_{\perp}^2 \times L_\|$, where the subscripts
denote the directions perpendicular and parallel to the background guide field $\vec{B_0}$,
respectively. We set $L_{\perp}=2\pi$, $L_\|/L_\perp=6$ and
$\vec{B_0}=5b_{rms}\vec{e_z}$, where $b_{\rm rms}$ is the root-mean-square average of the fluctuating magnetic field component.  The turbulence is driven at the largest scales by colliding Alfv\'en
modes~\footnote{Turbulence can also be driven by driving ${\bf v}$ or
  ${\bf b}$ fluctuations at large scales; this does not affect the
  inertial interval, see~\citep{mason_cb08}.}. We drive both Els\"asser
populations by applying statistically independent random forces
$\vec{f^+}$ and $\vec{f^-}$ in Fourier space at wave-numbers
$2\pi/L_{\perp} \leq k_{\perp} \leq 2 (2\pi/L_{\perp})$, $k_\| =
2\pi/L_\|$. The forces have no component along $z$ and are solenoidal
in the $xy$-plane.  All of the Fourier coefficients of $\vec{f^\pm}$ outside the above
range of wave-numbers are zero and inside that range are Gaussian
random numbers with amplitudes chosen so that $v_{rms}\sim 1$. The
individual random values are refreshed independently on average
approximately $10$ times per turnover of the large-scale eddies. The
variances $\sigma_{\pm}^2=\langle |\vec f^{\pm} |^2\rangle$ control
the average rates of energy injection into the $z^+$ and $z^-$
fields. In this work we consider the ``balanced'' case, that is we choose  $\sigma^+=\sigma^-$.  

A fully dealiased 3D pseudo-spectral algorithm is used to perform the spatial discretization on a grid with a resolution of $N_{\perp}^2\times N_\|$ mesh points (we typically take $N_{\perp}= N_\| = 512$  or 1024, see Table~\ref{table-sims}). 
We note that the domain is elongated in the direction of the guide field in order to accommodate the elongated wave-packets and to enable us to drive the turbulence in the strong regime while maintaining an inertial range that is as extended as possible \citep[see][]{perez_b10}.  This is a physical requirement that should be satisfied no matter what model system, full MHD or reduced MHD, is used for simulations.  
We also note that magnetic Reynolds number is defined by $R_m = b_{\rm rms} (L_{\perp} / 2 \pi) / \nu$ in our study.\footnote{In the case of reduced MHD though, when the $z^{\pm}_\|$ components are explicitly removed, the resulting system (\ref{rmhd-elsasser}) is invariant with respect to  simultaneous rescaling of the background field $B_0$ and the field-parallel spatial dimension of the system, if one neglects the dissipation terms. Therefore, one can rescale  the field-parallel box size to $L_\|=L_{\perp}$, that is, conduct the simulations in a cubic box, provided the backgrond field $B_0$ is rescaled accordingly.   We should note however that the dissipation terms in (\ref{rmhd-elsasser}) are not invariant and they should be changed accordingly under such rescaling.}   


\subsection{Numerical Procedures for Identifying and Characterizing Current Sheets}
\label{subsec-algorithms}

The question that we consider is: how can we unambiguously identify current sheets in a numerical simulation, where the thinness of developing current sheets is always limited by the grid resolution? To address this issue, we develop a concrete methodology and a specific algorithm for identifying structures in a large 3D simulation and for characterizing them quantitatively in terms of their geometrical properties and intensities.

Our algorithm is designed to be impartial to reconnection is the sense that it does not automatically associate the dissipation sites with magnetic X-points. Hence, the current sheets may or may not coincide with magnetic X-points. We consider the relationship to magnetic reconnection only after the current sheets have been fully characterized by their other properties. This differs from the approach of \cite{servidio2010}, where reconnection sites are detected first and the corresponding dissipation regions are studied afterwards.

\begin{figure}
   \includegraphics[width=8cm]{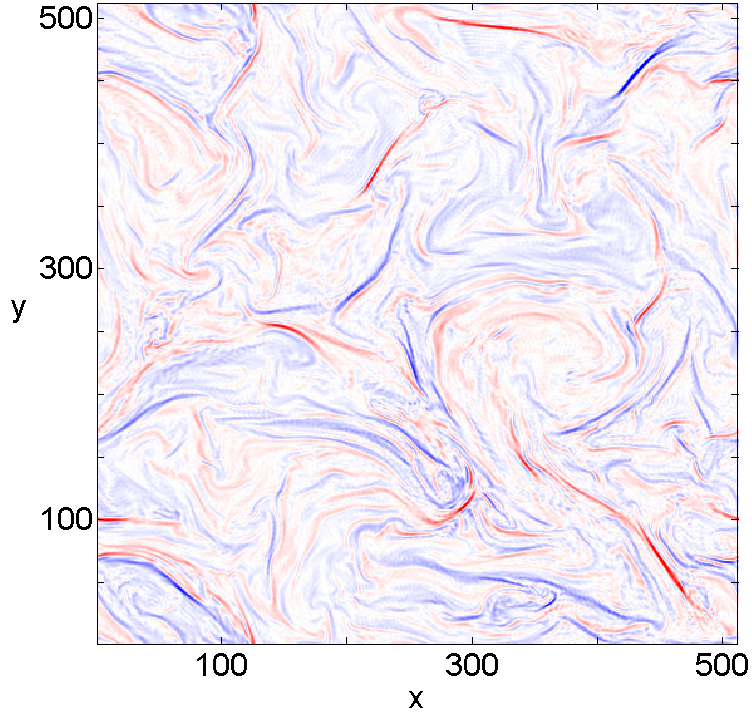}
   \centering
   \caption{\label{fig:profile} The current density in an $xy$-plane cross section of data. Red indicates negative current and blue indicates positive current. Current sheets protruding through the cross-section are clearly present.}
   \end{figure}

We implement the algorithm in IDL (Interactive Data Language). The algorithm can be applied either to 2D simulations or to 3D simulations with a strong magnetic guide field $B_0$ in the $z$-direction; here, we focus on the 3D case since it is more general. We assume that current sheets are predominantly aligned with the background guide magnetic field in order to simplify some parts of the algorithm (this assumption is supported by the results). The required input quantities are the magnetic flux function $\psi$, the current density $j = j_z = \nabla^2 \psi$, and the perpendicular ($xy$) magnetic field $\boldsymbol{b} = \hat{\boldsymbol{z}} \times \nabla \psi$. An illustration of the current density profile in an $xy$-plane cross section of our data is shown in Fig.~\ref{fig:profile}, where the presence of numerous current sheets penetrating the cross section is evident. 
  
   Before applying the algorithm, it is useful to interpolate the data to a higher-resolution grid by using a Fourier zero-padding and interpolation technique \citep{servidio2010}. This procedure consists of applying a fast Fourier transform (FFT) on the data, zero-padding the resulting set of coefficients to produce a larger array, and applying an inverse FFT to obtain the original points along with interpolated points in between. This interpolation scheme cannot be used on an entire 3D dataset because the computational costs are too great. However, it can be applied on any 2D cross-section of the data, from which we obtain a $4 N_x \times 4 N_y$ array in place of the original $N_x \times N_y$ array. This is useful because it improves the accuracy of several of our measurements, such as width and thickness, which are made in $xy$ planes. 

  We now describe our current-sheet identification algorithm. The first step is to locate the current sheets in a given simulation time snapshot. Since current sheets are characterized by extrema in the current density profile, the problem reduces to finding local maxima in the (magnitude of) current density. To achieve this, the algorithm scans through all points above a pre-specified threshold current density $j_{\rm thr}$ (which is several times larger than the globally averaged magnitude of current density) and selects those points that are local maxima within a surrounding cubic subarray of the data. Each subarray is centered at the candidate point and has size $(2 n + 1)^3$, where $n$ is a parameter. With larger $n$, the algorithm finds only the most dominant peaks. We choose $n = 4$ for our analysis, which allows well-resolved peaks to be detected. Our results do not change significantly if $n$ is adjusted by a factor of two or so. Every maximum is then identified with a current sheet that we label by index $i$, and the corresponding current density is referred to as the peak current density, $j_{\rm max,i}$.
  
  The second step is to identify the points belonging to each given current sheet. These are defined to be the points that collectively connect to the point of peak current density, with the condition that each point has a magnitude of current density greater than a minimal value, $j_{\rm min,i}$. For definiteness, we choose the current sheet boundary to be half of the peak current density, so $j_{\rm min,i} = j_{\rm max,i}/2$. The algorithm determines the current sheet points in the following way. First, it considers the points adjacent to the peak (four points for 2D analysis, six points for 3D analysis), and from these points adds the ones with current densities greater than $j_{\rm min,i}$ to a list. Then for each point now on the list, the algorithm adds to the list any unchecked adjacent points with current density above~$j_{\rm min,i}$. This procedure is repeated for subsequent points added to the list until no new points are found.
 Provided that $j_{\rm thr}$ is chosen to be high enough, the current sheets thus constructed are relatively isolated and sparse and do not form one globally percolating cluster. 
  
  There is, however, an ambiguity regarding how to treat current sheets that contain multiple peaks. If a particular current sheet has $j_{\rm min,i} > j_{\rm thr}$, then it is possible for a nearby peak to be associated with a second current sheet that contains points shared with the first current sheet. Whether or not such current sheets should be regarded as independent is unclear - this ambiguity is similar to the topography problem of objectively defining a mountain \citep{gerrard1990}. We choose to redefine current sheets with overlapping boundaries (by the criteria above) to be a single current sheet, with $j_{\rm max,i}$ taken from the most intense peak and $j_{\rm min,i}$ taken to be half of the smallest discernable peak within the composite current sheet. Therefore, a composite current sheet will have $j_{\rm min,i} < j_{\rm max,i}/2$. In general, the statistical properties and scalings of current sheets should be largely independent of how exactly the composite current sheets are defined.
  
\begin{figure}
   \includegraphics[width=8cm]{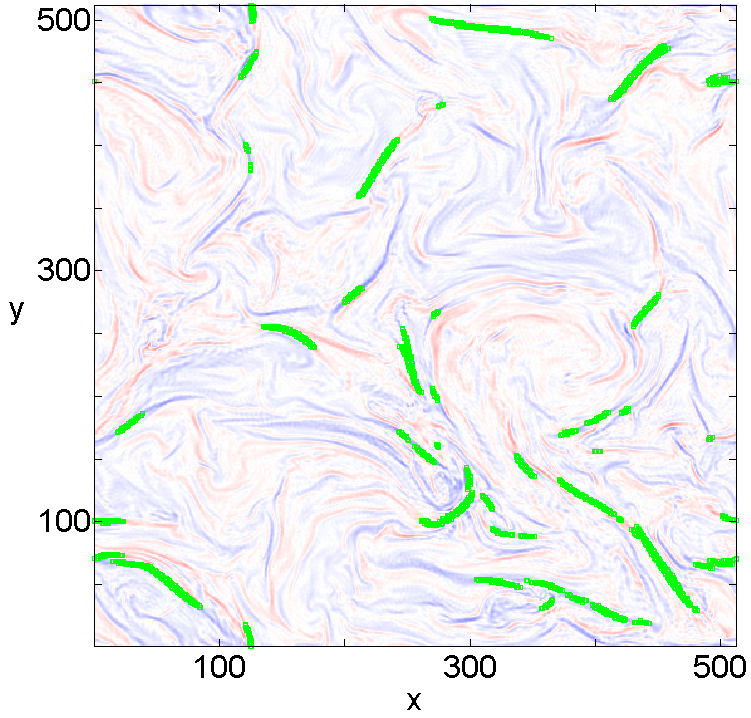}
   \centering
   \caption{\label{fig:area} Highlighted in green are the $xy$-plane cross sections of current sheets found by applying the algorithm to 3D current density data. The cross section of data is the same one as used in Fig.~\ref{fig:profile}. Note that the dominant structure is that of a current sheet.}
   \end{figure}

\begin{figure}
   \includegraphics[width=8cm]{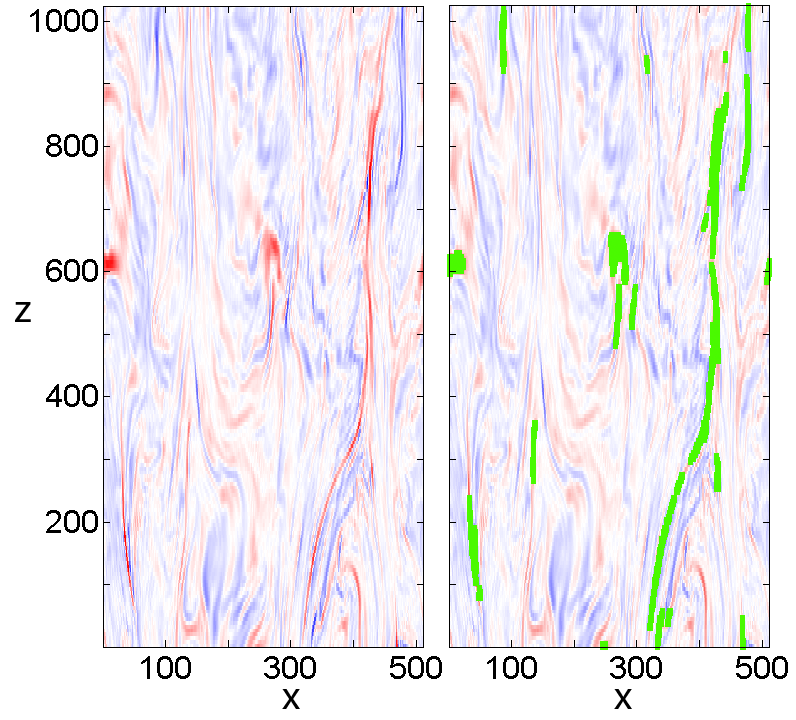}
   \centering
   \caption{\label{fig:parallel} On the left, an $xz$-plane cross section of the current density is shown, for $1/3$ of the simulation box height with actual aspect ratio. On the right, areas corresponding to current sheets, as found by applying the algorithm to the 3D current density data, are highlighted in green. Note that the current sheets are elongated in the direction parallel to the background guide field.}
   \end{figure}

  An illustration of the resulting cross sections for current sheets in our data is shown in Fig.~\ref{fig:area}.  Additionally, the current sheet cross sections in the $xz$ plane (parallel to the guide field) for $1/3$ of the simulation box height are shown in Fig.~\ref{fig:parallel}. It is clear that the dominant structure is indeed that of a current sheet --- a quasi-1D structure in a 2D slice or a quasi-2D ribbon in full~3D. Upon closer inspection, the actual 3D shape of a current sheet can often be complex. Common structural features include curvature, irregular boundaries, and strong asymmetry around the peak. This departure from the traditional, idealized Sweet-Parker picture of a straight and smooth current sheet can be attributed to turbulence. To some degree, the structure of an individual current sheet depends on the criteria in our definition; a current sheet that looks irregular when using the algorithm with one set of criteria may appear more regular when applying another set of criteria. However, the statistical conclusions of our study are not very sensitive to these criteria.
  
  With each individual current sheet now associated with a set of points in space, we can characterize their main physical properties quantitatively. In the following, we describe how we compute the ohmic energy dissipation, width, thickness, length, and upstream magnetic field vectors for a given current sheet. Note that we order the three current sheet dimensions by $\lambda < \xi < l$, where $\lambda$ is thickness, $\xi$ is width, and $l$ is length.

We start with $\cal{E}$, the total magnetic energy dissipated by the current sheet per unit time. Note that we consider only ohmic dissipation while ignoring viscous dissipation, which, in principle, can be of comparable magnitude in our simulation since we have $Pm = \nu/\eta= 1$.  Our diagnostic procedures can be generalized to include viscous dissipation rather straightforwardly and this is something we are planning to do in the near future. Since integration over the current sheet volume is numerically equivalent to a summation over all $N$ points, the energy dissipated becomes
  \begin{align}
  \cal{E} &= \iiint \eta j^2~{\textrm d}V = \sum_{k = 1}^N \eta j_k^2 h_x h_y h_z
  \label{eq:eneq1}
  \end{align} 
  where $j_k$ is the current density at the $k$th point of the current sheet, and $h_x$, $h_y$, and $h_z$ are the dimensions of each cell.


The measurements of thickness and width for each current sheet are made on the $xy$ plane that contains the point of peak current density of that current sheet.  Accordingly, it is useful to define the current sheet cross section to be the set of points in this plane that belong to the current sheet. 

\begin{figure*}
   \includegraphics[width=16cm]{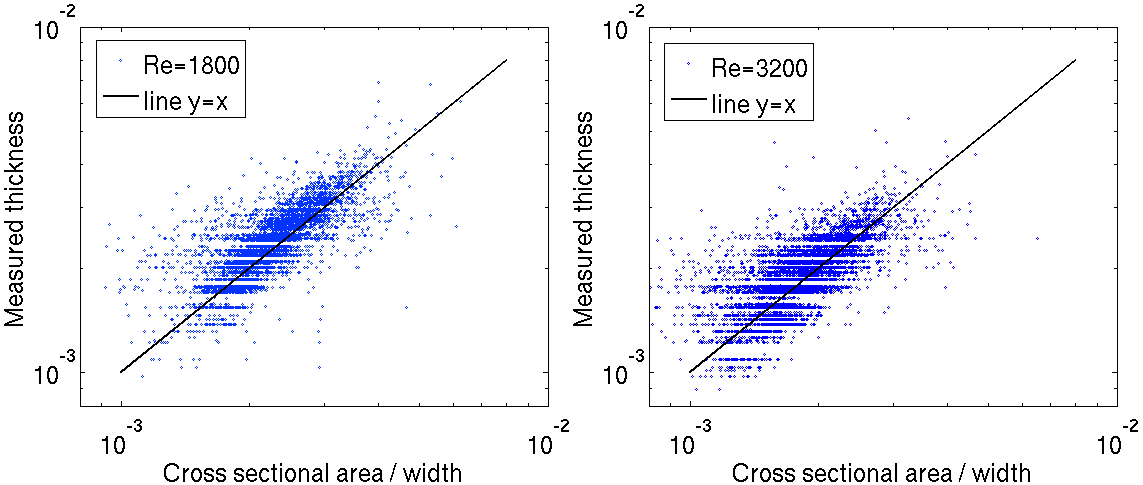}
   \centering
   \caption{\label{fig:thickness_est} A scatter plot comparing the thickness measured directly by the algorithm (on the $y$-axis) and the thickness estimated by dividing the current sheet cross sectional area by width (on the $x$-axis). The left panel shows the simulation with $R_m = 1800$ while the right panel shows the simulation with $R_m = 3200$. The significant amount of scatter between the two methods demonstrates the importance of choosing an appropriate method. Also, the increased amount of scatter in the $R_m = 3200$ case may suggest that the current sheets are not sufficiently resolved.}
   \end{figure*}

A relatively direct method for measuring current sheet thickness is as follows. We first determine the direction of most rapid descent from the peak current density. This is accomplished by numerically computing the Hessian matrix for the current density $j(x,y)$ at the peak,
  \begin{align}
  H &=
   \left(
   \begin{array}{cc}
   \partial_{xx} j &  \partial_{xy} j \\
   \partial_{yx} j &  \partial_{yy} j
   \end{array}
   \right)
   \end{align}
   and calculating the eigenvectors of $H$. We then find the distance from the peak in this direction at which the current density drops to that of the boundary, $j_{\rm min,i}$. We repeat this procedure in the opposite direction, and add the two distances to get the total sheet thickness~$\lambda$. We also employ an alternative way to estimate the thickness (similar to the one used by \cite{uritsky2010}) in which we divide the area of the current sheet cross section by its width (for which the measurement procedure is discussed in the next paragraph). A scatter plot comparing the thicknesses obtained by using the two different methods is shown in Fig.~\ref{fig:thickness_est} for simulations with magnetic Reynolds number $R_m = 1800$ and $R_m = 3200$ with resolution~$1024^3$. In general, we find a good agreement between these two independent methods. The increased amount of scatter for the $R_m = 3200$ case, which has thinner current sheets than $R_m = 1800$, suggests that those current sheets may not be sufficiently well resolved. Any measurement of thickness is intrinsically limited by the numerical resolution of the simulation. Indeed, the thickness obtained from the eigenvector method is quantized due to stepping across an integer number of cells of width $h$ (or $h/4$ when interpolation is used). The other thickness estimate is less affected by resolution, but loses information on the local thickness near the peak, the location of most interest.

Next, we define the width $\xi$ to be the largest distance between any pair of points in the $xy$ cross section of the current sheet --- this is very accurate unless the cross section is highly curved. It is found by iterating over all pairs of points in the cross section to obtain the maximum distance. It is worth noting that the width of the current sheet can be found alternatively by applying the eigenvector measurement procedure used for thickness, but using the other eigenvector of $H$ instead. However, this method is less robust since the boundary can be reached prematurely due to current sheet curvature. 

One may consider defining the length $l$ of the current sheet to be the maximum distance between any pair of points in the entire 3D current sheet ribbon. However, this is not useful in practice because it requires iterating over all pairs of points in the current sheet, which is computationally expensive for large 3D current sheets. Instead, we use the assumption that current sheets are aligned in the $z$-direction, and define the length to be the distance between the endpoints, which are defined to be the current sheet points with the maximum $z$ coordinate and the minimum $z$ coordinate. This estimate is accurate even for current sheets that are misaligned with respect to the $z$-direction, because we take into account the $x$ and $y$ coordinates of the endpoints.
   
 We also determine the magnetic field vector at key points of the current sheet. In particular, we measure the asymptotic (upstream) values of the $xy$ magnetic field on both sides of the current sheet. We define $B_{1,\parallel}$ and $B_{2,\parallel}$ to be the magnetic field components parallel to the current sheet at the two edges, which will be used later to compare to reconnection models. We first measure the magnetic field vectors $\boldsymbol{B}_1$ and $\boldsymbol{B}_2$ at the two edge points which we define to be the points used in the thickness measurement (see above). Then, $B_{1,\parallel}$ and $B_{2,\parallel}$ are found by projecting $\boldsymbol{B}_1$ and $\boldsymbol{B}_2$ onto the direction perpendicular to the eigenvector of $H$ in the $xy$ plane used to find the thickness. If $\boldsymbol{v}$ is the unit vector in the $xy$ plane that is orthogonal to the eigenvector of $H$, then
  \begin{align}
 B_{1,\parallel} &= \boldsymbol{B}_1 \cdot \boldsymbol{v} \nonumber \\
 B_{2,\parallel} &= \boldsymbol{B}_2 \cdot \boldsymbol{v} \label{eq:balg}
  \end{align}
   
  Finally, we consider the degree of association of the current sheets with magnetic reconnection events. For the purposes of this paper, we use magnetic X-points as a proxy for reconnection sites, although 3D reconnection is not always associated with null points \citep{priest1995,parnell2010}. Thus, we first detect X-points in the simulation, and then classify current sheets by whether or not they contain an X-point.

We note that X-points are equivalent to saddle points in the magnetic flux function. Therefore,  the problem in 2D reduces to detecting saddle points and determining whether they lie inside of current sheets. The 3D case is subtler because there is no analogue of a saddle point for a function in 3D space. However, we can again take advantage of the fact that reduced MHD is used, which allows us to consider saddle points in the magnetic flux function for every $xy$-plane that constitutes the volume. The set of these X-points then approximates X-lines where 3D reconnection takes place.

One way to find saddle points is by using the first and second derivative test. In this case, saddle points correspond to points where the two eigenvalues of the Hessian matrix for~$\psi(x,y)$, $H_\psi$, have opposite signs. We found, however, that this method did not work well for discrete numerical data. As a simple example, take a region of space where the function has integer values:
   \begin{center}
   2\ \ 2\ \ 0
   
   2\ \ 1\ \ 0
   
   0\ \ 0\ \ 2
   \end{center}
   The central point should be considered a saddle point because the function alternatively decreases in two directions and increases in two orthogonal directions. However, the slope does not change sign in the $x$-direction nor the $y$-direction, and so this saddle point would be missed by the first derivative test.
   
   We use an alternative method that always finds these saddle points, similar to one proposed by Kuijper \citep{kuijper2004}. The algorithm first considers the eight points surrounding the candidate point. If the function at these points, in clockwise order, rises above the candidate point's value twice and falls below the candidate point's value twice, then it follows that the candidate point is a saddle point. If the function rises and falls any other number of times, then it is not a saddle point.

\begin{figure}
   \includegraphics[width=4cm]{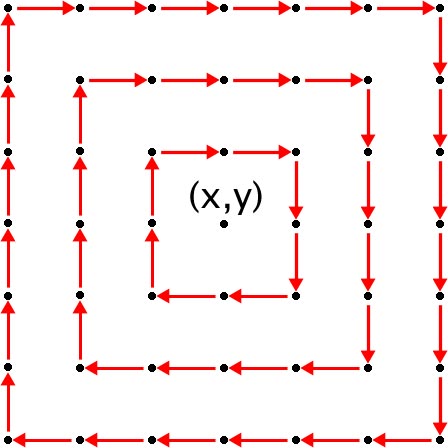}
   \centering
   \caption{\label{fig:1a} To determine whether a candidate point is a saddle point, consider the surrounding points along a loop and check whether those values twice rise above and twice fall below the value at the candidate point. To avoid detecting false saddle points, repeat the procedure for two larger loops shown.}
   \end{figure}
   
   A disadvantage of this procedure is that some false saddle points may be found if there is small-scale structure near the point. To avoid this, we repeat the procedure for two larger clockwise loops, having circumferences of sixteen and twenty-four points, respectively. This is shown schematically in Fig.~\ref{fig:1a}. If the pattern of rising twice and falling twice still holds for each of these loops, then we consider the candidate point to be a saddle point. After determining the location of all of the saddle points, it is straightforward to classify the current sheets by whether or not they contain an X-point, and are therefore associated with magnetic reconnection.
   
   This concludes the discussion of the algorithm. We have identified the locations and volumes of current sheets, measured several of their properties, determined the nearby magnetic field vectors, and classified them by whether or not they contain an X-point. The next section discusses the results that we obtain from applying the algorithm to our simulation data.


\section{Results}
\label{sec-results}

We analyze three different simulations of MHD turbulence. In Case 1, we use resolution of $512^3$ and magnetic Reynolds number $R_m = 1800$. In Case 2, we use resolution $1024^3$ and $R_m = 1800$. In Case 3, we use resolution $1024^3$ and $R_m = 3200$.  We analyze multiple time snapshots for each simulation (10 for Cases 1 and 2, and 7 for Case~3).  We find that the current sheets are best resolved in case~2, so we use data from that simulation for the majority of our analysis. The threshold current density for detection, $j_{\rm thr}$, differs for each case, as shown in Table~\ref{table-sims}.  We always choose $j_{\rm thr}$ much larger than the mean, $j_{avg} \approx 10$, and than the fiducial estimate~(\ref{eq-j_eta})  for the current density at the turbulent dissipation scale, $\tilde{j}_\eta \sim 30$. At the same time, we always choose $j_{\rm thr}$ to be significantly smaller than the global simulation box maximum, max($j$). For any given simulation, the lower limit for $j_{thr}$ is set by the value at which connected islands of high current density span the entire simulation box (similar to problems in percolation theory). By choosing the values of $j_{thr}$ shown in Table~\ref{table-sims}, the largest structures have a length comparable to the box size in the $z$ direction. Also shown in Table~\ref{table-sims} is the average number of current sheets detected per snapshot. The total number of current sheets in the sample are $N_{tot} = 7175$ for Case 1, $N_{tot} = 6616$ for Case 2, and $N_{tot} = 9343$ for Case~3.

\begin{table}[h!b!p!]
\caption{Simulation parameters \newline}
\centering
\begin{tabular}{|c|c|c|c|c|c|} 
	\hline
\hspace{0.5 mm} Case \hspace{0.5 mm}  & Resolution    &   \hspace{2 mm}$R_m$\hspace{2 mm} & \hspace{1 mm}$j_{thr}$\hspace{1 mm} & \hspace{1 mm}max($j$)\hspace{1 mm} & \hspace{0.5 mm}$N/$snapshot\hspace{0.5 mm}  \\
	\hline
1 & $512^3$ & 1800 & 100 & 437 & 718 \\
2 & $1024^3$ & 1800 & 130 & 656 & 662 \\
3 & $1024^3$ & 3200 & 160 & 837 & 1335 \\ 
	\hline
\end{tabular}
\label{table-sims}
\end{table}

We find that many current sheets (more than half of the total) do not contain X-points and, likewise, many X-points do not lie in strong current sheets. This lack of correlation suggests two things. Firstly, turbulence creates current sheets that are not located at sites of reconnection. Secondly, active reconnection does not occur at many X-points in the domain.
 
  We find, however, that the strongest current sheets tend to have X-points, and that X-point containing sheets exhibit different statistics in general. It is therefore reasonable to conclude that the strongest current sheets, and hence the most intense energy dissipation events are driven by reconnection, while the smaller ones are formed by turbulent fluctuations.
  
We now present the details of our quantitative statistical analysis. Note that all of the following probability distributions are normalized such that the integral over the given values is equal to unity. Also, distribution plots are shown on a log-log scale to highlight power law regions. The spatial scales are measured with respect to the size of the simulation box in the direction perpendicular to the guide field.

It is illuminating to first look at the mean values of the current sheet properties in the three simulations, even though the exact values depend on what detection threshold ($j_{\rm thr}$) is chosen. These properties are shown in Table~\ref{table-means}. It is immediately seen that the structures are indeed highly anisotropic, i.e. $\langle l \rangle \gg \langle \xi \rangle \gg \langle \lambda \rangle$. We also see that Case 1 and Case 2, which differ in resolution but have equal Reynolds numbers, exhibit similar length, width, and energy dissipation rates, despite having different threshold current densities. We see however that the mean thickness, $\langle \lambda \rangle$, changes significantly between the two cases, suggesting that it may be insufficiently resolved in Case 1. We also note that for all cases, the current sheets in total occupy less than $1\%$ of the system volume but account for roughly $25\%$ of the overall ohmic dissipation, showing that they are indeed a significant contribution to the total energy budget of the system.

\begin{table}[h!b!p!]
\caption{Mean values of current sheet parameters\newline}
\begin{tabular}{|c|c|c|c|c|c|} 
	\hline
\hspace{0.5 mm} Case \hspace{0.5 mm}  & \hspace{2 mm}$\langle l \rangle$\hspace{2 mm}    &   \hspace{4 mm}$\langle \xi \rangle$\hspace{4 mm} & \hspace{4 mm}$\langle \lambda \rangle$\hspace{4 mm} & \hspace{1 mm}$\langle j_{\rm max} \rangle$\hspace{1 mm} & \hspace{4 mm}$\langle \cal{E} \rangle$\hspace{4 mm}  \\
	\hline
1 & 0.39 & 0.049 & 0.0035 & 130 & $3.6\times10^{-4}$ \\
2 & 0.37 & 0.046 & 0.0024 & 182 & $3.5\times10^{-4}$ \\
3 & 0.29 & 0.033 & 0.0019 & 213 & $1.5\times10^{-4}$ \\ 
	\hline
\end{tabular}
\centering
\label{table-means}
\end{table}


\subsection{Statistical analysis}

\begin{figure}
 \includegraphics[width=8cm]{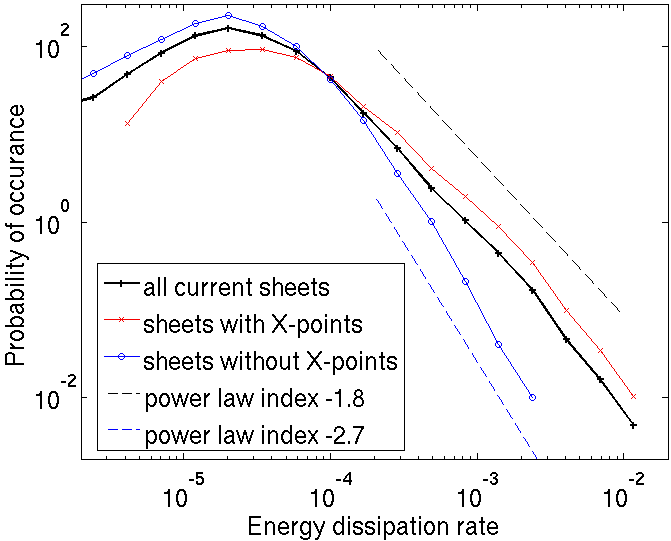}
   \centering
   \caption{\label{fig:edist} The probability distribution for current sheet ohmic energy dissipation rate, $\cal{E}$, measured using our algorithm and shown on a log-log plot. The distribution for all current sheets (in black) shows a power law tail with an index near $-1.8$. Also shown are distributions for two subpopulations of current sheets: those with X-points (in red) and those without X-points (in blue, with a power law index near $-2.7$). Current sheets without X-points dominate at low energies, while current sheets with X-points dominate at high energies.}
 \end{figure}

The probability distribution for energy dissipation rate integrated over a current sheet, $\cal{E}$ from Eq.~\ref{eq:eneq1}, is shown in Fig.~\ref{fig:edist}.  The distribution for all current sheets (black curve)  exhibits a power law tail (for ${\cal{E}} > 5 \times 10^{-5}$) with an index of $-1.8 \pm 0.1$. It is interesting to note that this power law is close to, but slightly harder than, the critical power law of index -2, indicating that the overall energy dissipation rate is dominated by large-$\cal{E}$ events.
Also shown in Fig.~\ref{fig:edist} are similar probability distributions for the sub-populations of current sheets with X-points and current sheets without X-points (in red and blue, respectively). The current sheets with X-points dominate at large energy dissipation rates while the sheets without X-points dominate at small energy dissipations. In fact, every single current sheet with ${\cal{E}} \gtrsim 2 \times 10^{-3}$ contains at least one X-point. Furthermore, approximately $84\%$ of the total energy dissipation of current sheets occurs in sheets with X-points, while the remaining $16\%$ occurs in sheets without X-points. The sheets without X-points exhibit a separate, steeper power law tail with an index close to $-2.7$.

\begin{figure}
\includegraphics[width=8cm]{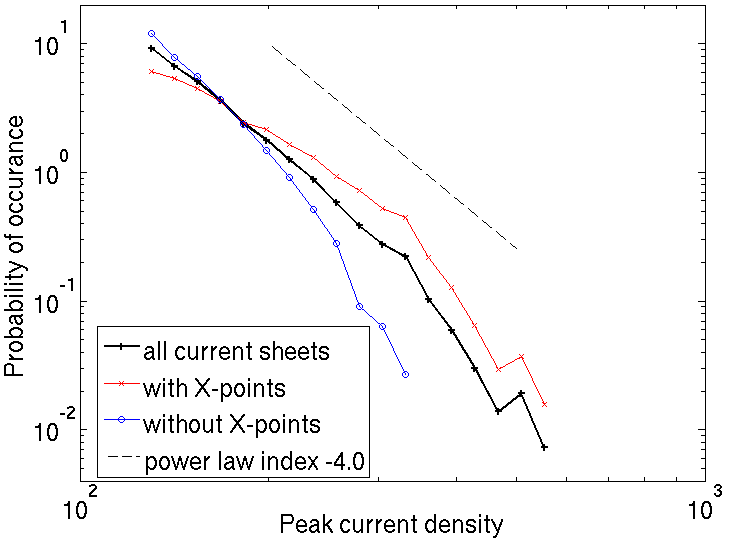}
   \centering
   \caption{\label{fig:jdist} The probability distribution for peak current density, $\j_{\rm max}$, in current sheets. Current sheets with X-points (in red) dominate over sheets without X-points (in blue) at large current densities, in agreement with the asymmetry in energy dissipation rate (see Fig.~\ref{fig:edist}).}
 \end{figure}

 The probability distribution for peak current density, $j_{\rm max}$, is shown in Fig.~\ref{fig:jdist}. In this case, there is a possible power law with index around $4$, but is difficult to measure because the range of current densities sampled is relatively narrow, less than an order of magnitude (due in part to our relatively high choice of current density threshold --- we only have a factor of 5 between the threshold and the global maximum current density). We find that the current sheets with large peak current densities tend to contain X-points, in agreement with the asymmetry in the distribution of $\cal{E}$ for the two subpopulations. This supports the picture that the strongest sheets tend to contain X-points.

\begin{figure*}
   \includegraphics[width=16cm]{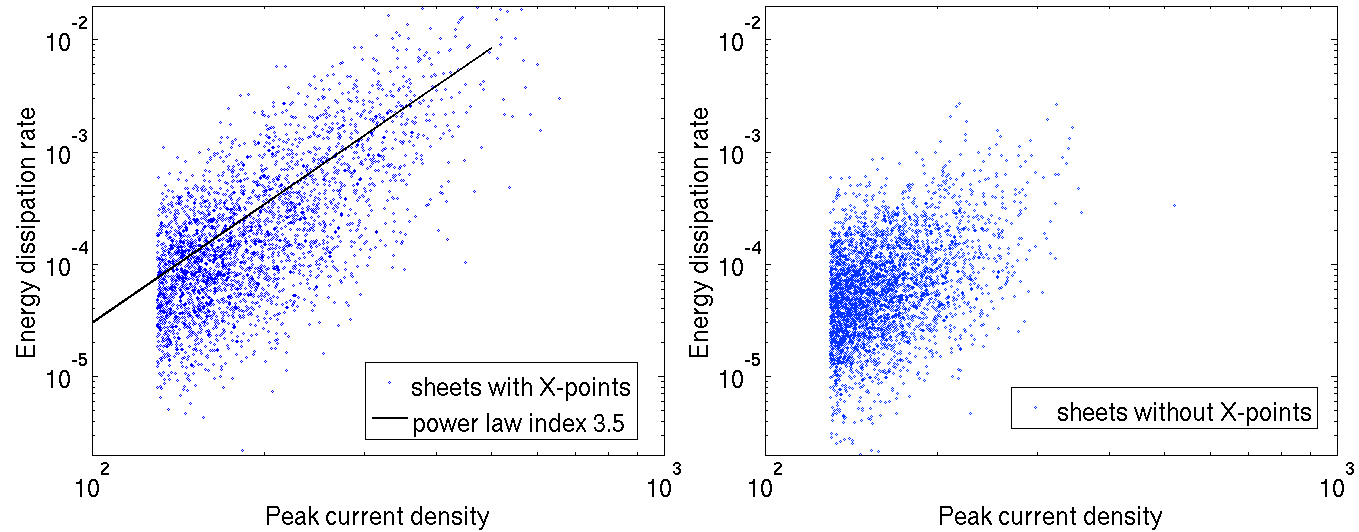}
   \centering
   \caption{\label{fig:strengths} As shown in the scatter plot on the left, there is a significant correlation between the ohmic energy dissipation rate and peak current density for current sheets that contain X-points, described approximately by a power law with index~$3.5$. However, as shown on the right, the correlation between the two quantities is much weaker for sheets that do not contain X-points.}
 \end{figure*}

The energy dissipation rate and peak current density are two different measures of a current sheet's strength, so we expect them to be correlated. Scatter plots of $\cal{E}$ versus $j_{\rm max}$ are shown in Fig.~\ref{fig:strengths} for the two populations of current sheets separately. We find that there is indeed a significant correlation between the $\cal{E}$ and $j_{\rm max}$ for current sheets with X-points, with a power law dependence ${\cal{E}} \propto j_{\rm max}^{3.5 \pm 0.5}$. However, we find that the correlation for the current sheets without X-points is much weaker.

\begin{figure*}
   \includegraphics[width=18cm]{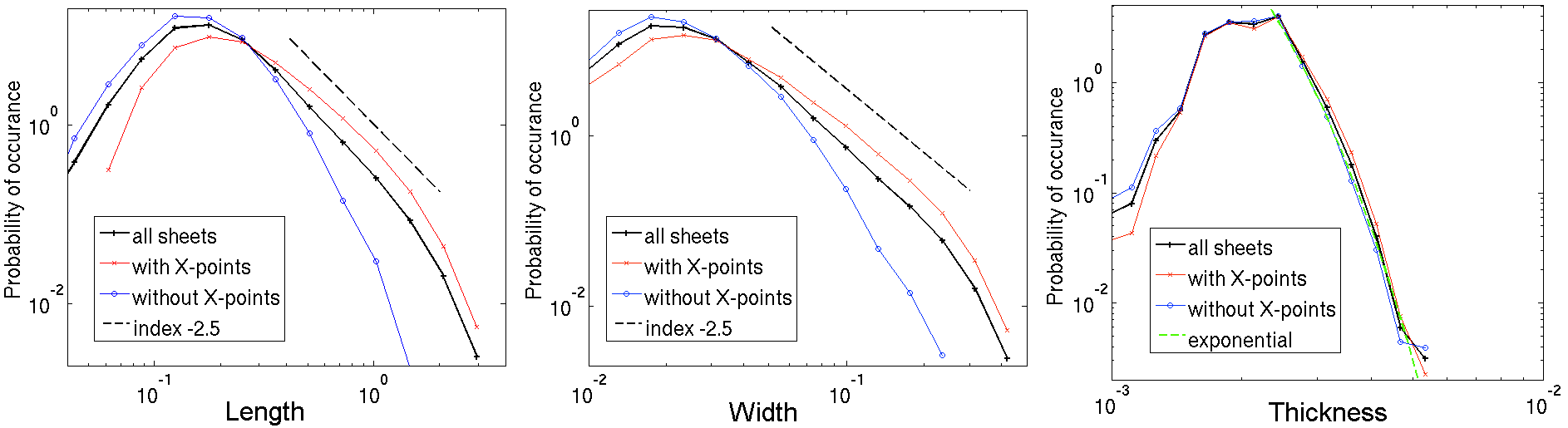}
   \centering
   \caption{\label{fig:dist_dims} The probability distributions for all current sheet dimensions. For the population of all current sheets (shown in black), the distributions for length $l$ and width $\xi$ have power law tails with indices near $-2.5$, while the thickness $\lambda$ has a much steeper tail (possibly an exponential, with a fit shown in green). The current sheets with X-points tend to be longer and wider (in red) than the sheets without X-points (in blue), but thickness is similarly distributed for all current sheets.}
 \end{figure*}

Now we discuss the statistical properties of the current sheet sizes. The probability distributions for length $l$, width~$\xi$, and thickness $\lambda$ are shown in Fig.~\ref{fig:dist_dims}. For the population of all current sheets, the distributions for length and width have power law tails, both with indices near $-2.5$. 
The thickness distribution declines much more rapidly at large~$\lambda$;  if fit to a power law, this distribution would have an index near $-9$, but the decline is so steep that is not clear that a power law is appropriate. An exponential decline appears to better match the data, with a fit proportional to $\exp{(-\lambda/(3.6\times10^{-4}))}$ shown.
The current sheets with X-points tend to be longer and wider than current sheets without X-points, while the distribution of thicknesses is similar for both types of current sheets.

\begin{figure*}
   \includegraphics[width=18cm]{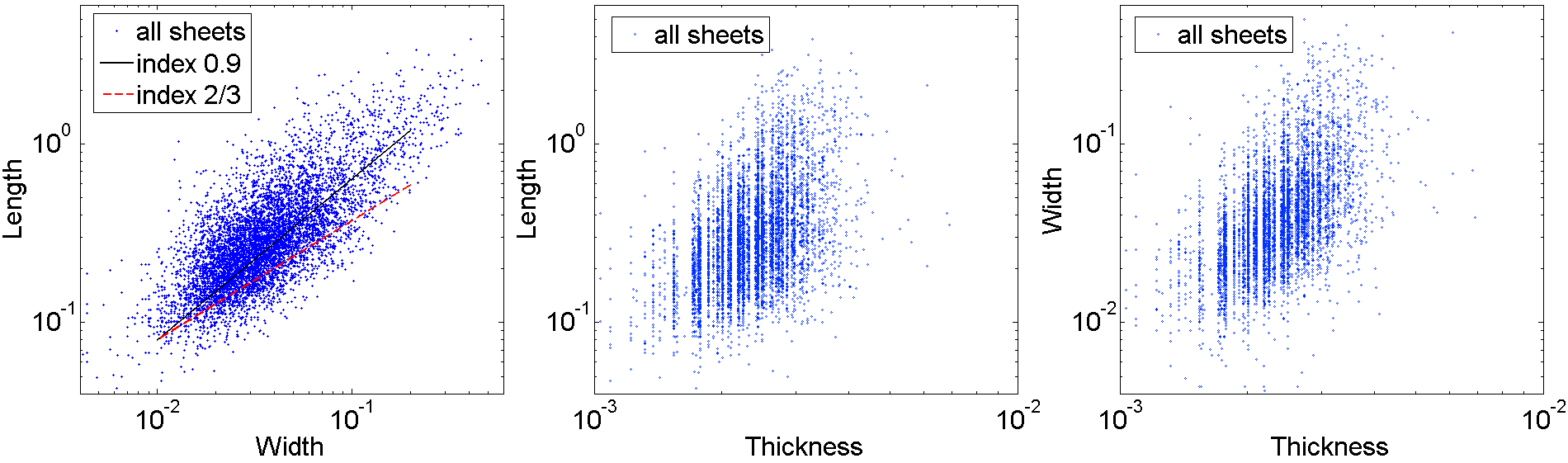}
   \centering
   \caption{\label{fig:dims} For the population of all current sheets, the length $l$ and width $\xi$ are strongly correlated with a power law fit of index $0.9 \pm 0.1$ (in black). This is somewhat steeper than the naively expected index of $2/3$ (in red). The current sheet thickness $\lambda$ is not strongly correlated with length or width.}
 \end{figure*}

Next we consider the cross-correlations of the current sheet dimensions.  The scatter plots in Fig.~\ref{fig:dims} show the correlation between all pairs of dimensions, for the population of all current sheets. The length and width are strongly correlated and have a nearly linear relationship, obeying a power law $l \propto \xi^{0.9 \pm 0.1}$. This is somewhat steeper than the index $2/3$ naively expected from MHD turbulence (from Eq.~\ref{lengtheq}). This disagreement, however, is not surprising since the naive expectation describes typical turbulent eddies, and there is no {\it a priori} reason why it should apply to the intense, strongly intermittent current sheets under investigation here.

For the population of all current sheets, we find that thickness is not strongly correlated with length or width. There appears to be a broad trend for thickness to increase with length and width, but it is dominated by scatter. Note also that thickness measurements cover a shorter range of values ($10^{-3} \lesssim \lambda \lesssim 5\times10^{-3}$) than the other two dimensions, making it difficult to discern any trend.

\begin{figure*}
   \includegraphics[width=15cm]{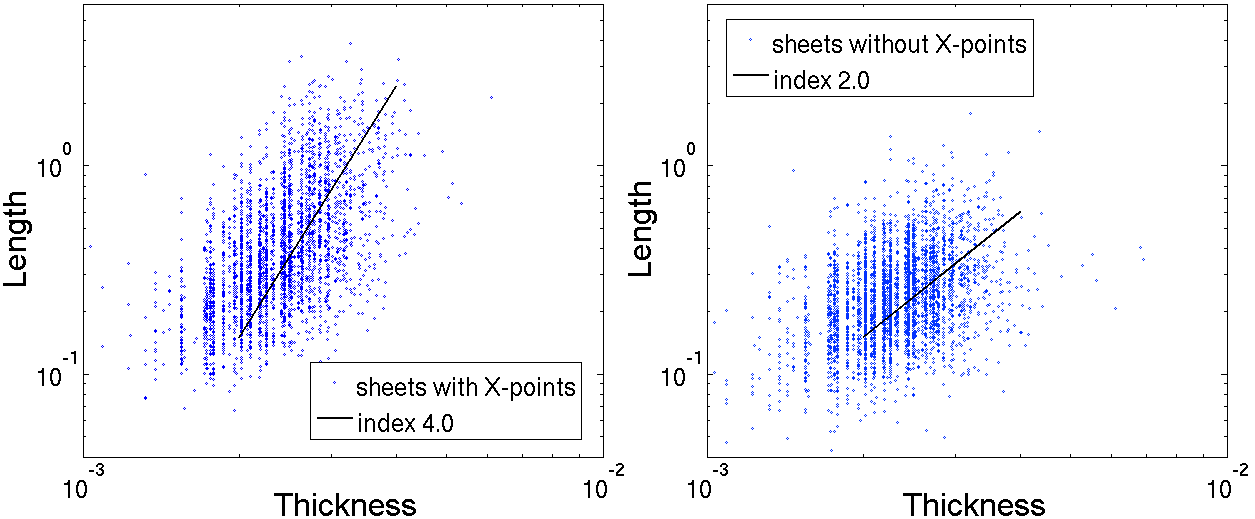}
   \centering
   \caption{\label{fig:lengthvthick} The current sheet length $l$ versus thickness $\lambda$ shown separately for sheets with X-points (left) and sheets without X-points (right). The sheets with X-points appear to have a stronger correlation between length and thickness, with a power law index near $4.0 \pm 1.0$. The length and thickness are not strongly correlated in sheets without X-points, with a possible power law index near $2.0 \pm 1.0$.} 
 \end{figure*}

\begin{figure*}
   \includegraphics[width=15cm]{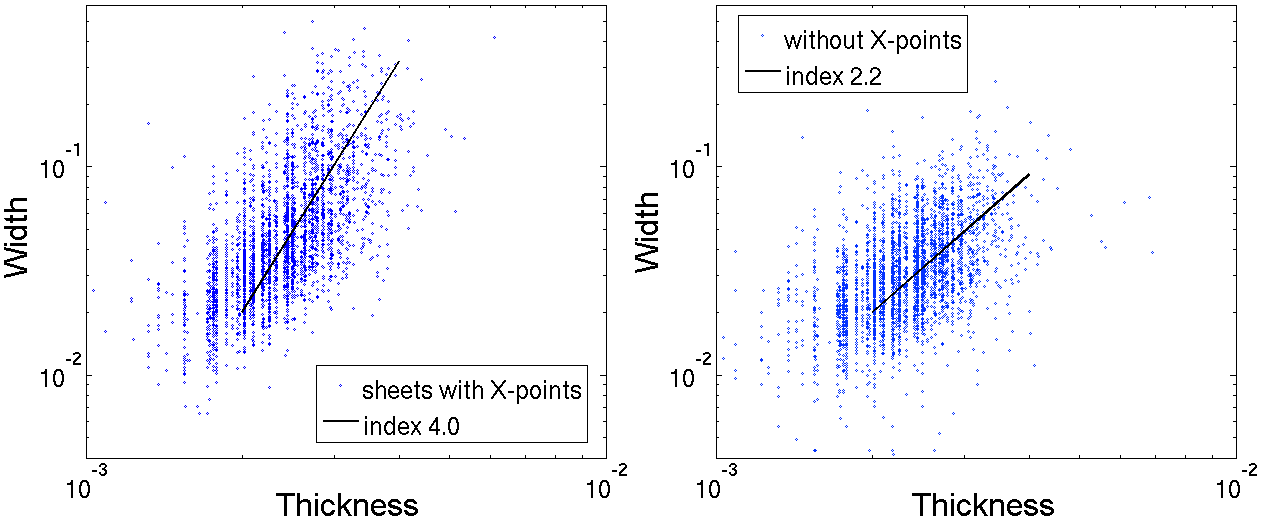}
   \centering
   \caption{\label{fig:widthvthick} The current sheet width versus thickness shown separately for sheets with X-points (left) and sheets without X-points (right). The sheets with X-points appear to have a stronger correlation between width and thickness, with a power law index near $4.0 \pm 1.0$ (similar to the length versus thickness fit, Fig.~\ref{fig:lengthvthick}). The width and thickness are not strongly correlated in sheets without X-points, with a possible power law index near $2.2 \pm 1.0$.} 
 \end{figure*}

The picture of the scaling of dimensions changes somewhat when we consider separately the sub-populations of current sheets with X-points and current sheets without X-points.  We find no significant difference between the two populations with regard to the scaling of length with width. However, the scaling with thickness does change. We find that both length and width are more strongly correlated with thickness for sheets that contain X-points, as shown in the left panels of Figs.~\ref{fig:lengthvthick} and~\ref{fig:widthvthick}.  The power-law fits for both of these correlations yield similar indices, $l \propto \lambda^{4.0 \pm 1.0}$ and $\xi \propto \lambda^{4.0 \pm 1.0}$.  Both of these scalings are much steeper than the scalings naively expected from MHD turbulence, which are $l \propto \lambda^{1/2}$ and $\xi \propto \lambda^{3/4}$ (see Eq.~\ref{lengtheq} and Eq.~\ref{widtheq}). This substantial disagreement between observations and the phenomenologically expected power laws could be explained by the fact that thicknesses lie within the dissipation range, where the MHD turbulence estimates do not apply. For current sheets without X-points (see the right panels of Figs.~\ref{fig:lengthvthick} and~\ref{fig:widthvthick}), we find that the correlations of lengths and widths with thickness are not as strong. A power-law fit gives $l \propto \lambda^{2.0 \pm 1.0}$ and $\xi \propto \lambda^{2.2 \pm 1.0}$, which is significantly shallower than the fit for the sheets with X-points. The difference in cross-correlations of the dimensions between sheets with X-points and sheets without X-points suggests that reconnection significantly affects the geometry of the current sheet.

\begin{figure*}
   \includegraphics[width=15cm]{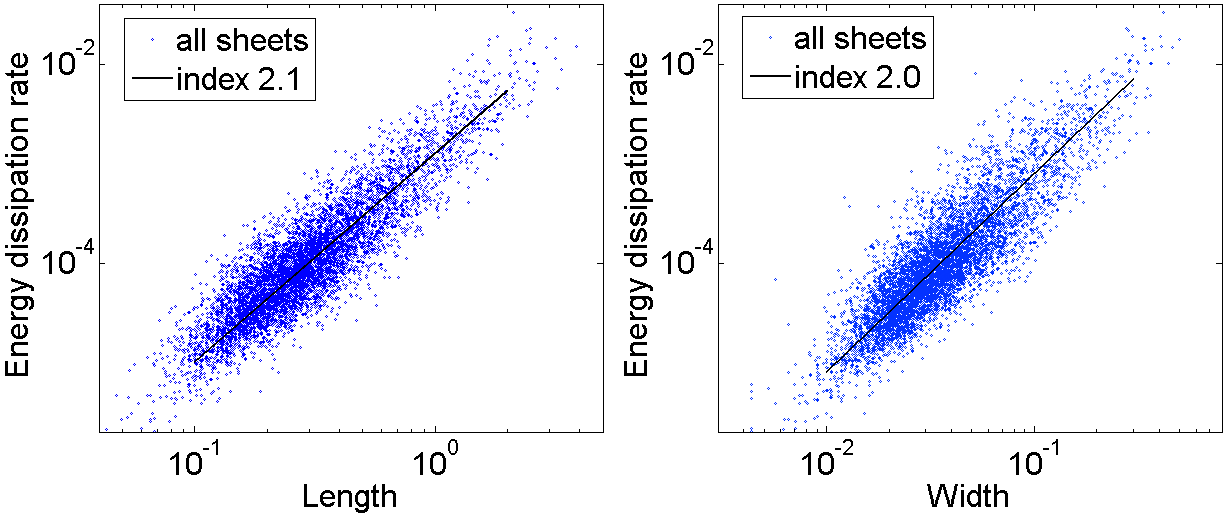}
   \centering
   \caption{\label{fig:ejvl} For all current sheets, there is a strong correlation between the energy dissipation rate and length (left), as well as for the energy dissipation rate and width (right). The relationships are approximately power laws, with indices $2.1 \pm 0.2$ and $2.0 \pm 0.2$ respectively.}
 \end{figure*}

\begin{figure}
   \includegraphics[width=8cm]{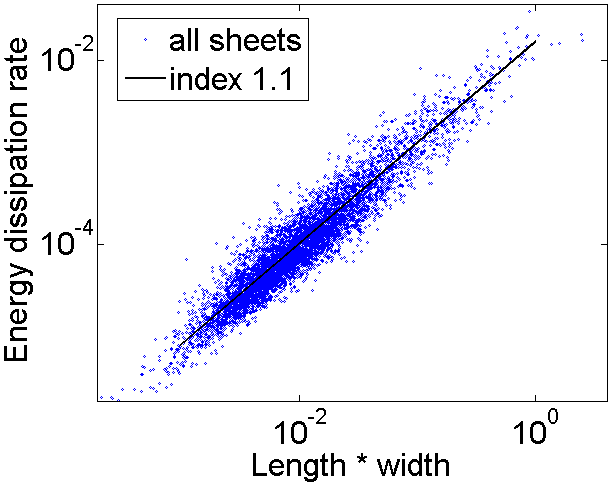}
   \centering
   \caption{\label{fig:eva} The energy dissipation rate and product of width and length are nearly proportional, with a good fit being a power law of index $1.1 \pm 0.1$.}
 \end{figure}

\begin{figure*}
   \includegraphics[width=15cm]{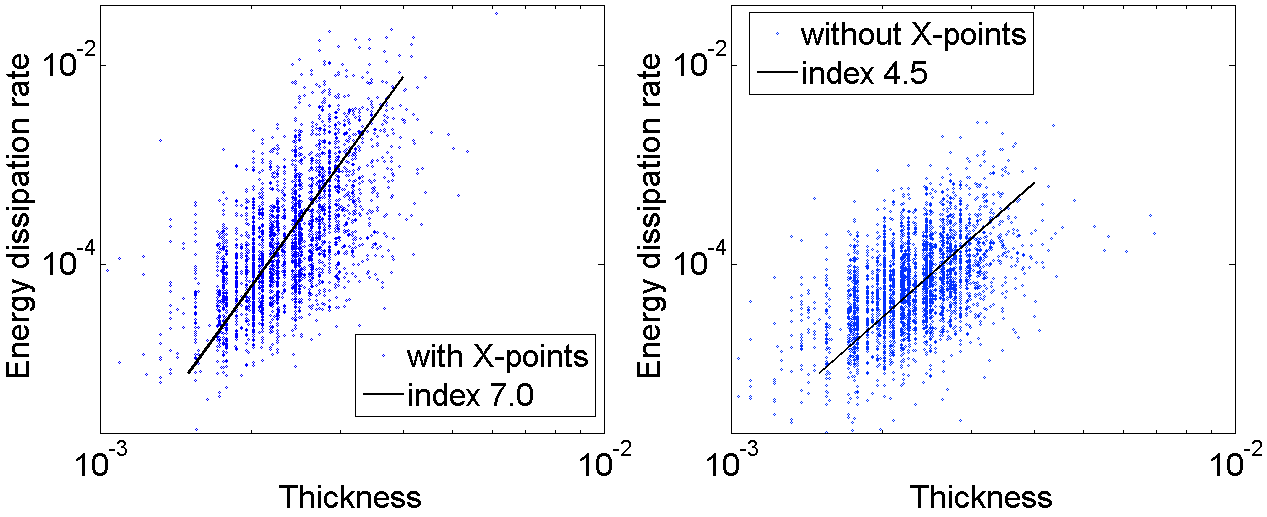}
   \centering
   \caption{\label{fig:evt} For current sheets with X-points, there is a weak correlation between the energy dissipation and thickness (left), fit to a power law with index $7.0 \pm 1.0$. For sheets without X-points, the correlation is similarly weak, fit to a power law with indices $4.5 \pm 1.5$.}
 \end{figure*}

The scalings of ohmic energy dissipation rate with the current sheet length and width are shown in Fig.~\ref{fig:ejvl}. The energy dissipation rate has a strong correlation with both these quantities. The scaling with length is approximately quadratic, $\mathcal{E} \propto l^{2.1 \pm 0.2}$, and the scaling for the width is similar, $\mathcal{E} \propto \xi^{2.0 \pm 0.2}$. All current sheets, regardless of whether or not they contain an X-point, exhibit this scaling. We also find that the energy dissipation rate is nearly proportional to the product of length and width, with a power law fit of $\mathcal{E} \propto (l \xi)^{1.1 \pm 0.1}$, shown in Fig.~\ref{fig:ejvl}. This suggests that neither thickness nor peak current density are important parameters in determining the total energy dissipation rate from a given current sheet.  Indeed, although we find a correlation between the thickness and energy dissipation rate, as shown in Fig.~\ref{fig:evt}, the correlation is relatively weak when compared to other correlations. This could also be expected because peak current density and thickness have a smaller spread of values than length and width.

\begin{figure*}
   \includegraphics[width=15cm]{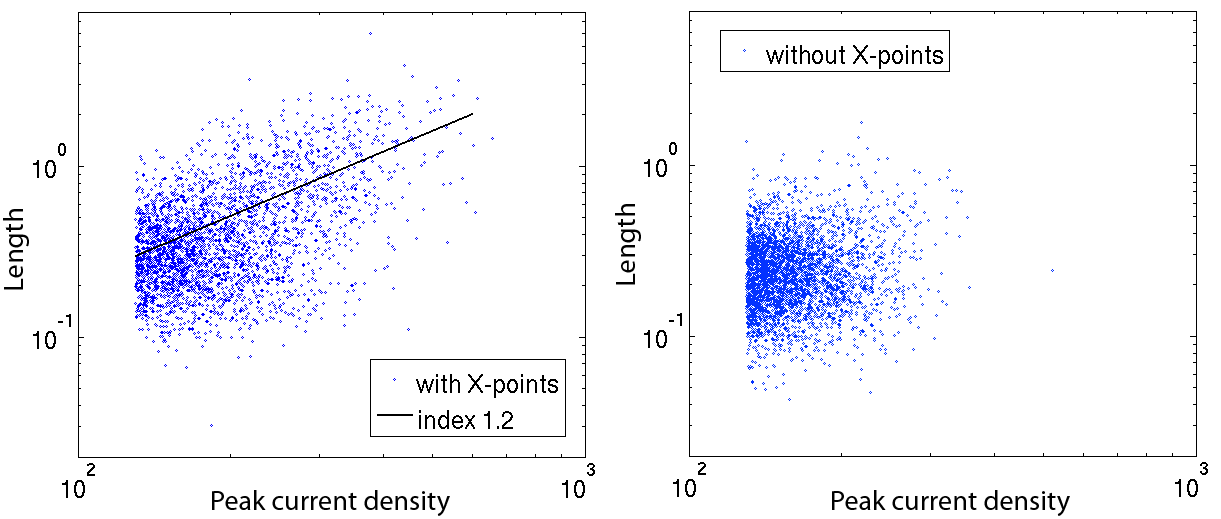}
   \centering
   \caption{\label{fig:lengthvpeak} Scatter plots of length versus peak current density, separately for sheets with X-points (left) and without X-points (right). Current sheets with X-points have a significant correlation with a power law index $1.2 \pm 0.2$, while there is no strong correlation for current sheets without X-points.}
 \end{figure*}

We now consider the scaling of length $l$ with peak current density $j_{\rm max}$, separately for current sheets with and without X-points. The scatter plot for this is shown in Fig.~\ref{fig:lengthvpeak}. There is no significant correlation between $j_{\rm max}$ and $l$ for current sheets without X-points. However, there is a moderate correlation between the two quantities for current sheets with X-points. A power law fit gives a relationship of $l \propto j_{\rm max}^{1.2 \pm 0.2}$. It is interesting that the scaling of energy dissipation rate with length is not significantly different for current sheets with and without X-points, but the scaling of peak current density with length does depend on the populations. This may indicate that the energy dissipation rate is essentially a geometric quantity, while the peak current density is more strongly influenced by dynamical effects such as reconnection and flow into an X-point.

\begin{figure*}
   \includegraphics[width=15cm]{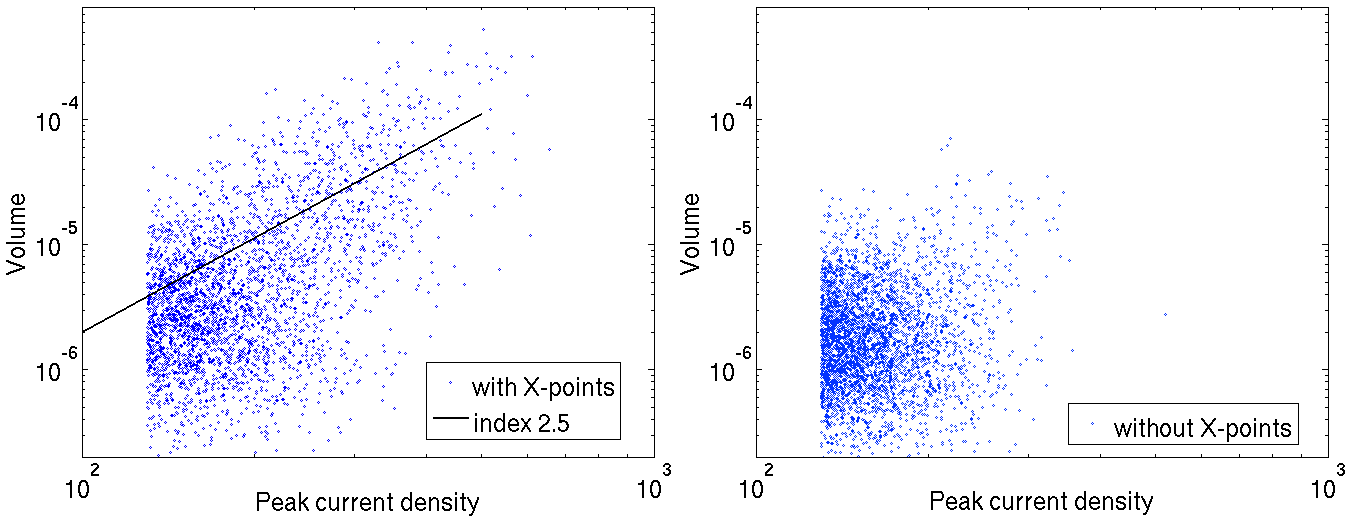}
   \centering
   \caption{\label{fig:jvv} There is a significant correlation between volume and peak current density for current sheets with X-points (left), although there is no strong correlation for sheets without X-points (right).}
 \end{figure*}

We find that there is a correlation between the intensity of the current sheet (measured by peak current density) and the volume of the current sheet, as shown in Fig.~\ref{fig:jvv}. In principle, this correlation must be taken into account when associating reconnection with current sheets that contain X-points, since a current sheet with a larger volume has a higher probability of containing an X-point purely by chance. This effect in itself could produce the X-point asymmetry in the distributions of $\cal{E}$ or $j_{\rm max}$. To address this, we determine whether the probability of randomly containing an X-point is important by using the following method. We redo the analysis with uniformly scattered random points replacing the actual locations of X-points. Then we classify the current sheets by whether or not they contain these random points. If current sheets with random points reproduce the same asymmetry in the energy dissipation distribution as observed for current sheets with X-points, then we can conclude that the current sheets are not correlated with X-points. Upon performing this test, we found that the current sheets with random points cannot account for the distributions observed by current sheets with X-points. Therefore, strong current sheets are indeed most likely to form at the locations of X-points, although this does not change the fact that a significant number of intense current sheets exist without X-points.


\subsection{Comparison to Sweet-Parker model}

The final part of our analysis is a test of the applicability of the Sweet-Parker \citep{sweet1958, parker1957} model of reconnection, which predicts the reconnection rate and current sheet thickness at a site of magnetic reconnection, to X-point-containing intense current sheets produced in MHD turbulence.  Although other resistive-MHD reconnection models exist, we consider the Sweet-Parker model for the following reasons.  First, numerical simulations of reconnection in resistive MHD have shown that Petschek's \citep{petschek1964} model fails while the Sweet-Parker model provides a robust solution, at least in regime of modest Lundquist numbers~\citep{biskamp1986, uzdensky2000}. Second, while it is now well established that Sweet-Parker current sheets with high Lundquist numbers ($S > S_c \sim 10^4$) become unstable to the plasmoid (secondary tearing) instability \citep{loureiro2007}, which completely disrupts them \citep{samtaney2009, bhattacharjee2009, huang2010, loureiro2012}, the current sheets in our simulations are never long enough for their corresponding Lundquist numbers to exceed the instability threshold~$S_c$. Therefore, plasmoid-dominated reconnection is not relevant in our simulations. Finally, \cite{servidio2009} have shown that the Sweet-Parker model can locally describe reconnecting current sheets in 2D MHD turbulence, making it reasonable to hypothesize that 3D current sheets behave similarly.

The Sweet-Parker model \citep{sweet1958, parker1957} assumes a 2D current sheet of roughly uniform (or slowly varying) current density in a high-aspect-ratio, approximately rectangular region called the reconnection layer.  The reconnection rate and the reconnection layer thickness can then be derived entirely from conservation laws \cite[for a derivation, see, for example,][] {biskamp2003}. The classic Sweet-Parker model predicts that for symmetric reconnection with a corresponding current sheet of width $\xi$ and an upstream reconnecting magnetic field of strength~$B$, the layer thickness should be
\begin{align}
\lambda_{SP} \propto \left(\frac{\xi \eta}{B}\right)^{1/2}  = \xi\, S^{-1/2} \, , 
\label{eq:sweetparkerb}
\end{align}
where $S \equiv \xi B /\eta$ is the Lundquist number for this layer and $B$ is taken in units of the Alfv\'en velocity. 
A slightly modified form of this equation is obtained by estimating $B$ in terms of the peak current density in the current sheet, $j_{max} \approx B/\lambda$, which gives
\begin{align}
\lambda_{SP} \propto \left(\frac{\xi \eta}{j_{max}}\right)^{1/3} \label{eq:sweetparker}
\end{align}

\begin{figure*}
   \includegraphics[width=15cm]{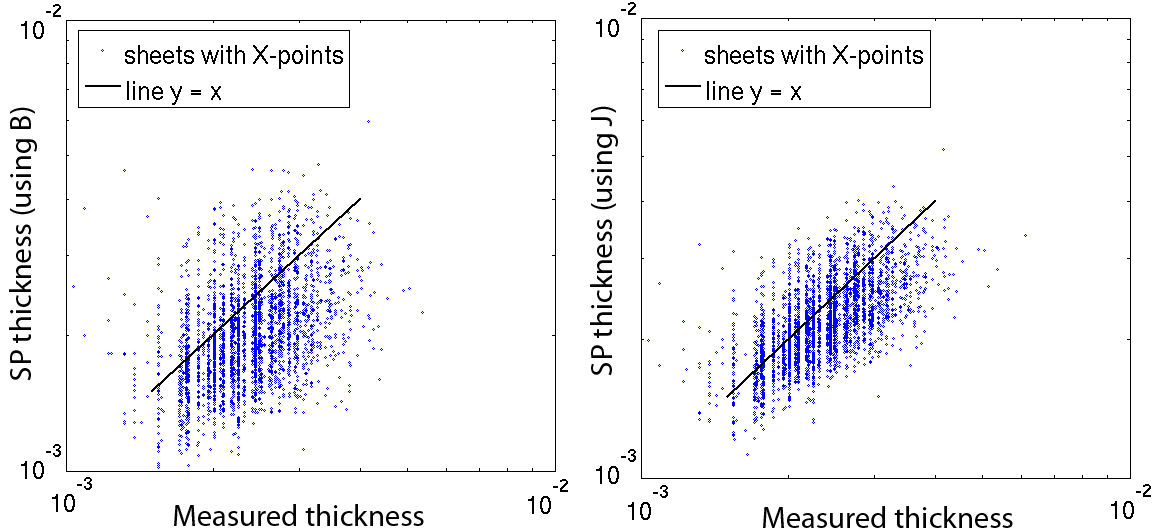}
   \centering
   \caption{\label{fig:sweetparker} Comparison of measured current sheet thicknesses to Sweet-Parker theory predicted thicknesses. In the left panel, the Sweet-Parker thickness is computed using measured width and upstream magnetic field $B = (B_1 - B_2)/2$ (see Eq.~\ref{eq:sweetparkerb}). In this case, there is a large amount of scatter about the expected values. In the right panel, the Sweet-Parker thickness is computed using measured width and peak current density (see Eq.~\ref{eq:sweetparker}). In this case, there is better agreement between the measurements and predictions.}
 \end{figure*}

 We test the theoretical prediction of the Sweet-Parker model for our population of X-point-containing current sheets.  We use measurements of width, $\xi$, and the mean increment in upstream magnetic field, $B = |B_1 - B_2|/2$, where the fields $B_1$ and $B_2$ are the in-plane cross-section-parallel components of the magnetic field at the two edges of the current sheet, as explained in Subsection~\ref{subsec-algorithms} (see Eq.~\ref{eq:balg}).  We use these measurements to compute the expected thickness $\lambda_{SP}$ from Eq.~\ref{eq:sweetparkerb}.  We then compare $\lambda_{SP}$ to the actual (directly measured) thickness $\lambda$ for the same current sheet.  A scatter plot for this comparison, using current sheets from the Case 2 simulation, is shown in the left panel of Fig.~\ref{fig:sweetparker}.  We observe a large amount of scatter around the expected trend.  However, if we compute the Sweet-Parker thickness from Eq.~\ref{eq:sweetparker}, using the measured peak current density $j_{max}$ instead of the magnetic field jump, then we find a better agreement and a tighter correlation between the measurements and the predictions, shown in the right panel of Fig.~\ref{fig:sweetparker}. This suggests that the Sweet-Parker model actually provides a reasonable physical picture for what happens in these intense current sheets. In the other two simulations (Cases 1 and~3) the agreement is not as strong, which suggests that the current sheets in those cases may not be sufficiently resolved.
 
\begin{figure}
   \includegraphics[width=8cm]{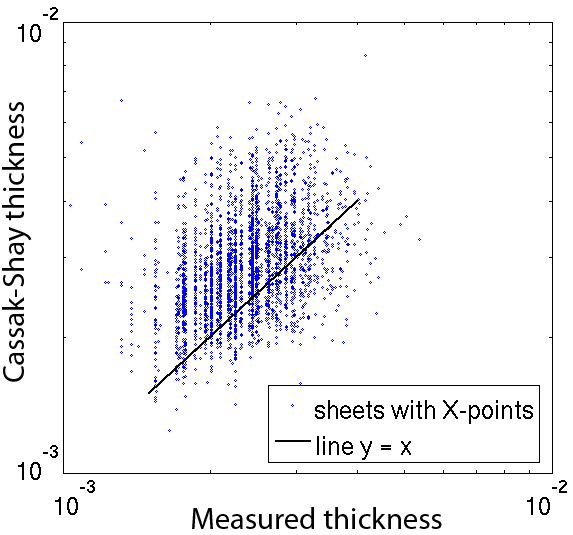}
   \centering
   \caption{\label{fig:cassakshay} Comparison of measured current sheet thicknesses to the predicted thickness for asymmetric current sheets in the Cassak-Shay model (see Eq.~\ref{eq:cassakshay}). The results are similar to the Sweet-Parker theory using an upstream magnetic field of $B = (B_1 - B_2)/2$ (see Fig.~\ref{fig:sweetparker}).}
 \end{figure}

We note, however, that the original Sweet-Parker theory is intended for describing only the so-called symmetric reconnection case, $B_1 = - B_2$.  Current sheets that are formed in MHD turbulence have no a priori reasons to be symmetric, however.  Therefore, we also consider the generalization of the Sweet-Parker model to asymmetric ($B_1 \neq - B_2$) current sheets, developed by~\cite{cassak2007}. In this case, the X-point does not coincide with the center of the current sheet. For upstream parallel magnetic fields $B_1$ and $B_2$, \cite{cassak2007} find 
\begin{align}
  \lambda_{CS} \propto \frac{(\eta\xi)^{1/2}}{(B_1 B_2)^{1/4}} \left[ \left( \frac{B_1}{B_2} \right)^{1/2} + \left( \frac{B_2}{B_1} \right)^{1/2} \right]
  \label{eq:cassakshay}
  \end{align} 
We test this prediction in a similar way to the symmetric Sweet-Parker model. For the upstream values of $B_1$ and $B_2$, we take the measured fields and subtract off the same component of the magnetic field at the current density peak. This removes the effect of the large scale field and improves agreement. 
The results are very similar to our comparison with the Sweet-Parker model using $B = (B_1 - B_2)/2$, as  shown in Fig.~\ref{fig:cassakshay}. 

There are a number of reasons why one may expect a certain degree of disagreement and a lack of tight correlation between the measured thickness and the thickness predicted by the Sweet-Parker and Cassak-Shay theories.  
In particular, turbulence distorts the current sheets significantly, resulting in violation of some of the assumptions  about the current sheet geometry that go into these simplified theoretical models.  
In addition, these models are stationary, whereas the current-sheet structures developing in our simulations are transient and highly non-stationary.


\section{Discussion and Conclusions}
\label{sec-conclusions}

This completes our statistical analysis of dissipative currents sheets in driven, incompressible, reduced-MHD turbulence.  We have arrived at a detailed picture of the typical current sheets that arise spontaneously from nonlinear interactions. This analysis has led to the following conclusions.

1) In MHD turbulence, a significant fraction of all ohmic dissipation takes place in highly intermittent, relatively extended structures --- in current sheets that are larger and more intense than the typical dissipative scale naively predicted by MHD turbulence theory. These structures are relatively rare and sparse but account for a significant fraction of all dissipation. For example, in Case 2, more than $25\%$ of all dissipation takes place in current sheets with central current peak density higher than approximately $8$ times the global root-mean-square current density; these current sheets occupy less than $1\%$ of the volume. 

2) Even though there is a natural tendency to automatically associate current sheets with reconnection
, we find that this is not necessarily correct.  In particular, we find that about $55\%$ of the intense current sheets, accounting for approximately $16\%$ of the overall energy dissipation from current sheets, do not contain magnetic X-points (whose presence is seen as a hallmark of reconnection). Likewise, we find that there is a large number of X-points in the simulation domain at any given time that are not associated with strong dissipation sites. Thus, one must be careful in postulating a direct correspondence between reconnection sites and energy dissipation sites in MHD turbulence. 

3) At the same time, however, we find that the most intense current sheets tend to be reconnecting ones. Thus, whereas only about half of the current sheets are found to contain X-points, their fraction increases as one considers more intense dissipation events. For example, the probability of a current sheet to contain an X-point increases to $90 \%$ when its corresponding peak current density becomes roughly half of the global maximum. 
Furthermore, these current sheets appear to exhibit different statistical properties in general; for example, their properties are more strongly correlated with thickness and peak current density. These current sheets appear to conform, on average, to the Sweet-Parker reconnection theory.



\subsection{Implications}

Our results have potential implications for turbulent energy dissipation and plasma heating in many space and astrophysical environments, including the solar wind, accretion disks and flows, the interstellar medium (ISM), and the hot coronal gas in galaxy clusters.

In particular, these findings may strongly influence our understanding of thermodynamics of space and astrophysical plasmas where the temperature structure is controlled by a balance between heating by turbulent energy dissipation and radiative cooling, such as accretion disks and the ISM.  
Quite often one treats these environments as having a well-defined quasi-uniform (e.g., slowly varying in space) temperature, determined by balancing the {\it volume-average} dissipation rate with the average radiative cooling rate.  However, this assumption is valid only if either the heating is uniform or if the heat released at localized dissipation sites is able to distribute itself more or less uniformly throughout the system, e.g., via thermal conduction, on time scales faster than the radiative cooling time. 
If, as we in fact found in the present study, turbulent heating is highly localized and intermittent, then, depending on the radiative cooling function and on the efficiency of thermal conduction, it is in principle possible that the plasma does not have a quasi-uniform, smooth temperature field.  Instead, the system is characterized in this case by an essentially inhomogeneous and intermittent thermal structure, with constantly appearing and disappearing relatively sparse hot regions that are quickly heated  by dissipation in current (and vorticity) sheets and then quickly cooled by radiation cooling,  before they can share their heat with nearby regions via thermal conduction.

The present work also has potentially important implications for situations in which the most intense current layers, responsible for a significant fraction of the overall dissipation, undergo a transition to fast collisionless reconnection. This may enhance the intermittent character of energy dissipation even further.

Our results may also have implications for the solar wind. Magnetic discontinuities, characterized by rapid spatial variation in the magnetic field vector, have been measured $\text{\it{in-situ}}$ by spacecraft in the solar wind. The observed probability distribution of discontinuities in the solar wind have been closely reproduced by simulations of MHD turbulence \citep{greco2010, zhdankin2012}.  In our present study, we found that reconnecting current sheets are indeed marked by a strong discontinuity in the magnetic field, and are reasonably consistent with Sweet-Parker current sheets. This may suggest that the solar wind contains intermittent current sheets that dissipate a significant fraction of the magnetic energy. If a more direct association between magnetic discontinuities (possibly measured by using several different methods) and properties of the corresponding current sheet can be established, then the solar wind turbulence and energy dissipation can be understood in greater detail.


\subsection{Comparison to similar studies}

In this subsection, we compare our results to similar statistical studies of intermittent structures in MHD turbulence. Our study was preceded by and partially motivated by the statistical analysis of \cite{servidio2009, servidio2010}, performed on magnetic reconnection sites in simulations of 2D MHD turbulence. 
Their algorithm is, broadly, as follows: the first step is to detect X-points, and the second step is to study the surrounding dissipation region (reverse of our procedure). They estimate the two dimensions of the diffusion region by using the eigenvectors and eigenvalues of the Hessian of magnetic potential. Hence, it is a local estimate that may miss information about the current sheet shape away from the X-point. In order to be accurate, this approach requires a very well resolved current density peak near the X-point. Our technique for measuring dimensions is more direct and is applicable for a wider variety of current-sheet shapes (for example, highly curved sheets). Although the \cite{servidio2009, servidio2010} algorithm is an excellent approach for studying the reconnection process itself, an important limitation is that it ignores current sheets not directly associated with an X-point. Our procedure extends this idea, making it applicable to dissipation sites not involved with reconnection.

\cite{servidio2009, servidio2010} find that if the reconnection rate at an X-point is greater than a fixed threshold, then generically the X-point is surrounded by a current sheet.  For X-points with a low reconnection rate, they find no clear scaling properties of the diffusion region parameters, suggesting that current sheets are not discernable. For X-points with a high reconnection rate, they find scalings consistent with the \cite{cassak2007} predictions for asymmetric Sweet-Parker reconnection. The upstream magnetic fields $B_1$ and $B_2$ and the current sheet length (the width in our 3D terminology) are used to test the theory, similar to our analysis. They obtain a stronger agreement with the Sweet-Parker model than we do, which may be due to using 2D simulations. Another advantage of their study is that the dissipation range in 2D is significantly better resolved than is possible in 3D simulations.


\cite{uritsky2010} performed a statistical analysis that more directly studies intermittent structures in MHD turbulence.  Although the analysis in our present paper is in many respects similar to that in their paper, there are also several important differences.  First, whereas their paper considers decaying MHD turbulence, our present study focuses on driven MHD turbulence.  Our approach then has the advantage of being able to study turbulence in a statistical steady state; our simulations can in principle run indefinitely, which allows us to collect the statistical data using many snapshots and thus obtain a much larger statistical sample. 
The second difference is that, whereas our study focuses on reduced-MHD turbulence, valid in the presence of a strong guide field, their investigation does not have this limitation; it uses the full incompressible MHD equations and does not assume a strong guide field.  In this sense, methodologically, their procedures are more universally applicable than ours. 
Thirdly, their cluster algorithm detects not only current sheets but also vorticity structures, which enables them to study the statistical properties of viscous dissipation, in addition to ohmic dissipation of interest to us here.  In this sense their analysis is more comprehensive, although one does not expect  significant differences between the two dissipation mechanisms in the case of MHD turbulence, as their study in fact confirms.  Adding viscous dissipation diagnostics to our study is a straightforward generalization of our algorithm and is something we plan to implement in the near future. 

Another important difference in terms of diagnostics is that \cite{uritsky2010} do not report the statistics of the peak current density and do not explicitly mention the value of the current density threshold used in their study; thus, it is not clear to what extent their conclusions apply to the most intense dissipative structures (which constitute the main focus of our present study).  Also, whereas their algorithm measures several geometric properties of the dissipative structures (such as the length), the current sheet thicknesses 
are not measured directly.  Instead, they are estimated as volume divided by surface area, which is, however, a valid approximation as long as the structures are sheet-like.  Finally, \cite{uritsky2010} do not explore or discuss any association of turbulent dissipative structures with magnetic reconnection sites, which is the main goal of our paper.   In particular, they do not distinguish between reconnecting and non-reconnecting current sheets, as marked by the presence or absence of magnetic X-points. 

Despite these differences, there are also many important similarities and points of comparison between \cite{uritsky2010} and our study.  We both find and analyze power-law distributions and scaling relationships for many of the measured current-sheet properties.  For some of them we can make a direct comparison and we find that our results are generally in reasonable agreement.  For example, in our study we find that the current-sheet length and width are distributed with a power-law of index of about -2.5, which is not too different from their value of $-\tau_L \simeq - 2.2 \pm 0.22$ for their Run~I (see their Table~II). Likewise, the ohmic dissipation rate integrated over a current sheet has a power-law distribution with an index of about -1.8 in our study, which is comparable to $-1.44 \pm 0.06$ in their study. We find that the ohmic dissipation rate scales with length as a power law with index 2.1, while they find an index $2.44 \pm 0.10$.  In addition, we both find that the distribution of current-sheet thicknesses $\lambda$ is significantly different from that of most other quantities --- it declines very steeply at large~$\lambda$, which may be more consistent with some sort of exponential cut-off rather than with a genuine power law.  This reflects the fact that current sheet thicknesses vary across a much narrower dynamic range of values and hence do not correlate very strongly with, e.g., the current-sheet length.


\section{Future directions}
\label{sec-future}

The study reported in this paper should be seen as only a first step in a much broader research program devoted to understanding the interaction between turbulence and reconnection in magnetized plasmas.  
The statistical analysis methods we developed can be further improved and extended in the following ways: 

(1) Supplementing our existing magnetic-field/current diagnostics with an analogous velocity/vorticity diagnostics, which will enable us to study the statistical properties of viscous dissipation in addition to ohmic dissipation. 

(2) Adding a temporal dimension to our study, i.e., tracking current sheets not only in space but also in time, determining their lifetime in addition to their spatial extent.  This should allow us to treat the dissipative spatio-temporal structures as events (flares or flashes) rather than just spatial structures and hence investigate the statistics of their total (time-integrated) energy release, lifetimes,  and related quantities.  
By tracking the evolution of individual current sheets in time, we should be able to determine whether they coalesce and decay, and to assess the validity of a stationary model for reconnection.  Our present simulation data sets are not resolved well enough in time to perform such an investigation. In principle, however, this can be done in the future by either having a very high cadence of saving numerical snapshots to the disk or by embedding the current-sheet diagnostic procedures directly into an MHD code itself and thus doing the analysis simultaneously with the simulation. 

(3) The present statistical analysis can also be improved upon by using data from better resolved simulations, which should increase the accuracy of measurements and reveal stronger trends over a wider sampling range.  In particular, this could strengthen the correlations associated with peak current density and thickness.  Indeed, it is known that accurate measurements of higher-order statistics require a very well resolved dissipation range, which is particularly important for numerical studies of reconnection \citep[see, e.g.,][]{wan2010}.  It is difficult to determine whether the current sheets in our present study are adequately resolved until improved simulations become available.  Also, we plan to perform a sequence of studies with several different values of the resistivity and viscosity. This will enable to see which statistical properties investigated in the present paper exhibit universal behavior, i.e., are independent of the magnetic Reynolds number, and also to see whether and how they depend on the magnetic Prandtl number.

Finally, we expect that the diagnostic methods and algorithms developed in the present study will find useful applications (perhaps with modest modifications) in various specific areas of modern heliospheric physics and astrophysics, such as: \\
-  other types of MHD turbulence --- most notably, turbulence driven by the magneto-rotational instability (MRI) in astrophysical accretion disks; \\
-  chaotic magnetic fields in magnetically-dominated environments, e.g., in the solar corona in the context of the coronal heating problem (and also in coronae of accretion disks, e.g., in black-hole systems); \\
-  relativistic turbulence \citep{zrake2013}; \\ 
-  non-MHD, purely hydrodynamic turbulence --- namely, quasi-1D dissipative structures (vortices) in incompressible turbulence and quasi-2D dissipative structures (shocks) in compressible supersonic turbulence, e.g., in molecular clouds \citep{boldyrev2002}.


\begin{acknowledgments}


This work was supported by National Science Foundation's 2010 Research Experiences for Undergraduates (REU) Program at the University of Colorado, by the NSF-sponsored Center for Magnetic Self-Organization in Laboratory and Astrophysical Plasmas at the University of Wisconsin - Madison, and by the US DoE awards DE-FG02-07ER54932, DE-SC0003888, and DE-SC0001794.  D.U. and S.B. thank the Aspen Center for Physics (supported by NSF Grant 1066293) where part of this work was performed in the Summer of 2006. 

\end{acknowledgments}


\bibliographystyle{apj}


\end{document}